\documentclass[sigplan,10pt,screen=true]{./templates/acmart}
\renewcommand\footnotetextcopyrightpermission[1]{} 
\setcopyright{none}
\usepackage{color, xcolor}
\usepackage{xurl}
\usepackage{xspace}
\usepackage{algorithm}
\usepackage[noend]{algpseudocode}
\usepackage{textcomp,upquote,listings, paralist}
\usepackage{enumitem}

\usepackage{amsmath}
\usepackage[font=small,labelfont={small,bf}]{caption}
\usepackage[font=small,labelfont={small,bf}]{subcaption}
\usepackage{hyperref}
\usepackage{tabularx}
\usepackage{comment}
\usepackage{booktabs}
\usepackage{balance}
\usepackage{multirow}
\usepackage{makecell}
\usepackage{amssymb}
\usepackage{pifont}
\usepackage{tikz}
\usepackage{multirow}
\usepackage{fancyhdr}

\newcommand*\circled[1]{\tikz[baseline=(char.base)]{
            \node[shape=circle,draw,inner sep=0.1pt] (char) {#1};}}

\definecolor{codegreen}{rgb}{0,0.6,0}
\lstdefinestyle{myStyle}{
    commentstyle=\color{codegreen},
    basicstyle=\footnotesize\ttfamily,
    keywordstyle=\color{blue},
    basewidth=0.5em,
    breakatwhitespace=false,
    stringstyle=\color{red},
    frame=tb,
    framerule=0.5pt,
    breaklines=true,
    keepspaces=true,
    numbers=left,
    numbersep=5pt,
    showspaces=false,          
    showstringspaces=false,
    showtabs=false,     
    tabsize=2,
    language=Python,
    morekeywords={onSelectPath,onChunkSize,onPacingChunk,onTxRtxChunk,onRxRtxChunk,onRxChunk,onRxACK,onRxCredit},
    deletendkeywords={bool},
}

\lstdefinestyle{APIDX_C_STYLE}{
    language=C,                            
    basicstyle=\fontfamily{pcr}\selectfont\small,
    keywordstyle=\color[HTML]{1922fb}\bfseries,     
    commentstyle=\color[HTML]{007020}\bfseries,     
    tabsize=4,                             
    breaklines=true,                       
    breakatwhitespace=true,                
    escapeinside={\%*}{*)},                
    morekeywords={uint32_t, uint64_t}
}
\usepackage[colorinlistoftodos,prependcaption,textsize=tiny]{todonotes}
\usepackage{verbatim} 
\newcommand{\hide}[1]{}

\newcommand{\ie}{i.e.\xspace}
\newcommand{\eg}{e.g.\xspace}

\newcommand{\para}{\noindent\textbf}

\newcommand{\paraspace}[1]{\vspace{0.03in}}

\newcommand{\presec}{\vspace{-0.10in}}
\newcommand{\postsec}{\vspace{-0.00in}}
\newcommand{\presub}{\vspace{-0.10in}}
\newcommand{\postsub}{\vspace{-0.02in}}
\newcommand{\presubsub}{\vspace{-0.05in}}

\newcommand{\prefig}{\vspace{-0.00in}}
\newcommand{\postfig}{\vspace{-0.10in}}
\newcommand{\postfigdouble}{\vspace{-0.25in}}
\newcommand{\postfigcaption}{\vspace{-0.20in}}

\setlist[itemize]{noitemsep, topsep=0pt, leftmargin=*}
\setlist[enumerate]{noitemsep, topsep=0pt, leftmargin=*}

\newcommand{\sysname}{UCCL\xspace}
\newcommand{\cx}{CX\_IB\xspace}
\newcommand{\cxeth}{CX\_ETH\xspace}
\newcommand{\alltoall}{all-to-all\xspace}
\newcommand{\allreduce}{allreduce\xspace}
\newcommand{\allgather}{allgather\xspace}
\newcommand{\reducescatter}{reduce-scatter\xspace}
\newcommand{\mallreduce}{multi-allreduce\xspace}
\newcommand{\malltoall}{multi-all-to-all\xspace}
\newcommand{\mallgather}{multi-allgather\xspace}
\newcommand{\mreducescatter}{multi-reduce-scatter\xspace}

\newcommand{\ualltoall}{All-to-All\xspace}
\newcommand{\uallreduce}{Allreduce\xspace}
\newcommand{\uallgather}{Allgather\xspace}
\newcommand{\ureducescatter}{Reduce-Scatter\xspace}
\newcommand{\umallreduce}{Multi-Allreduce\xspace}
\newcommand{\umalltoall}{Multi-All-to-All\xspace}
\newcommand{\umallgather}{Multi-Allgather\xspace}
\newcommand{\umreducescatter}{Multi-Reduce-Scatter\xspace}

\begin{document}

\date{} 



\begin{abstract}
Fast-evolving machine learning (ML) workloads have increasing requirements for networking. 
However, host network transport on RDMA NICs is hard to evolve, causing problems for ML workloads. 
For example, single-path RDMA traffic is prone to flow collisions that severely degrade collective communication performance. 
We present \sysname, an extensible software transport layer to evolve GPU networking. 
\sysname decouples the data path and control path of existing RDMA NICs and efficiently runs the control-path transport on host CPUs. 
This software extensibility brings in transport innovations that cannot be achieved in hardware for ML workloads, \eg, a multipath transport to resolve flow collisions. 
ML collectives atop \sysname achieve up to 4.5$\times$ higher performance compared to existing RDMA NICs. 
\end{abstract}

\newcommand{\titlespace}{\hspace{10pt}}
\title{An Extensible Software Transport Layer for\\ GPU Networking}
\author{\large 
\vspace{-0.05in}
Yang Zhou$^{1,2}$\footnotemark[1] \titlespace Zhongjie Chen$^3$\footnotemark[1] \titlespace Ziming Mao$^1$ \titlespace ChonLam Lao$^4$ \titlespace Shuo Yang$^1$ \titlespace \\ 
Pravein Govindan Kannan$^5$ \titlespace Jiaqi Gao$^6$ \titlespace Yilong Zhao$^1$ \titlespace Yongji Wu$^1$ \titlespace Kaichao You$^{1,3}$ \\
\titlespace Fengyuan Ren$^3$ \titlespace Zhiying Xu$^7$\footnotemark[2] \titlespace Costin Raiciu$^8$ \titlespace Ion Stoica$^1$\\
\vspace{0.05in}
\it $^1$UC Berkeley \titlespace $^2$UC Davis \titlespace $^3$Tsinghua University \titlespace $^4$Harvard University \titlespace $^5$IBM Research \\ 
$^6$Unaffiliated \titlespace $^7$Amazon Web Services \titlespace $^8$University Pollitehnica of Bucharest \& Broadcom}

\maketitle
\pagestyle{plain} 

\renewcommand{\thefootnote}{\fnsymbol{footnote}}
\footnotetext[1]{Equal contribution.}
\footnotetext[2]{This work does not relate to the position at Amazon.}
\renewcommand{\thefootnote}{\arabic{footnote}}

\sloppy
\presec
\section{Introduction}
\label{sec:intro}

Machine learning (ML) workloads and their requirements for networking are evolving rapidly. 
Less than ten years ago, deep neural networks only had millions of parameters, and were trained atop hundreds of CPUs/GPUs with parameter servers or \allreduce collective communication~\cite{ps_nips}. 
After five years, large language models (LLMs) began to surge with billions of parameters, and were trained atop thousands of more powerful GPUs with multi-level parallelism and diverse collectives like \allreduce, \allgather, and \reducescatter~\cite{llama3_tech, google_pathway}. 
In the recent two years, large-scale LLM serving has become the norm; prefill-decode disaggregation~\cite{distserve}, as an efficient serving technique, requires fast peer-to-peer communication. 
This year, serving Mixture-of-Experts (MoE) models like DeepSeek-V3~\cite{deepseek_v3} became very popular, featuring challenging \alltoall communication among hundreds of GPUs. 

However, networking techniques especially the host network transport on RDMA NICs are hard to adapt and evolve to better suit the needs of ML workloads.
Essentially, hardware changes are time-consuming and take much longer time than software changes. 
This can lead to a mismatch between the application needs and existing hardware optimizations, which often translates into poor performance. 
For example, Meta has reported that DCQCN~\cite{dcqcn}---a popular congestion control (CC) algorithm in datacenters supported by RDMA NICs---does not work well for LLM training workloads with low flow entropy and high traffic burstiness~\cite{llm_meta}. 
As a result, Meta decided to disable the CC support in NICs and instead implement traffic scheduling at the application layer. Similarly, DeepSeek disabled the CC when running large-scale \alltoall for serving MoE models~\cite{deepep}. 
However, running a large-scale RDMA network without CC is brittle, as it can lead to deadlocks, head-of-line blocking, and pervasive congestion~\cite{pfc_deadlocks, rdma_at_scale, irn, dcqcn, azure_rdma_storage, ali_rdma_storage}. 

In another example, Alibaba has observed severe performance degradation for collective communication during LLM training. This was due to the high level of flow collisions, which in turn was caused by the RDMA NICs supporting only single-flow/path per connection~\cite{llm_alibaba}. To avoid this problem, Alibaba has redesigned the network topology for LLM training using a rail-optimized dual-plane architecture. However, such a redesign is costly to build and maintain. As we will show in \S\ref{ssec:multipath}, a software-only solution implementing multipath at the application level would have avoided such topology changes.

In \S\ref{ssec:motivate_extend}, we discuss four more examples where ML workloads require adapting RDMA hardware support, including network incast in MoE serving, semi-reliable gradient transmission, efficient loss recovery, and avoiding vendor lock-in. Furthermore, the existing hardware-baked host network transport layer makes it difficult to productize new research proposals~\cite{mlt, strack, smartt, eqds} to improve the performance of ML workloads in real-world scenarios. 

These examples point to \emph{network extensibility} as one of the key challenges in datacenter networks. In this paper, we focus on addressing this challenge in the context of RDMA-enabled hosts, by proposing and implementing a software-only extensible transport layer called \sysname (Ultra-CCL). 
This software-only approach makes it much easier to efficiently support new ML workloads, \eg, CC algorithms that can better support ML traffic than existing DCQCN, and multipathing to mitigate flow collisions. 
To implement this approach, we need to address two challenges: (1) How to decouple the data and control paths for existing RDMA NICs? The data path handles network data transfers for the GPU, while the control path manages transport control decisions like CC, packet reliability, and multipath load balancing (LB). Only if the control path is decoupled from the data path can we implement it on the CPU. 
(2) How can we achieve hardware-level performance with the control path running on the CPU? This is challenging given the inter-GPU-server high bandwidth, which can exceed 3.2 Tbps~\cite{aws_p5e, azure_h100}.

To address the first challenge, \sysname repurposes the features of existing RDMA NICs to achieve efficient data and control path separation. 
In particular, \sysname leverages RDMA Unreliable Connection (UC) to bypass hardware-baked CC and packet reliability logic, and uses RDMA immediate data to pass transport control states between sender and receiver CPUs.
For some NICs that do not support UC, such as AWS EFA~\cite{efa}, \sysname leverages the scatter-gather feature over RDMA Unreliable Datagram (UD) to achieve separation. 

To address the second challenge, \sysname leverages an array of techniques tailored to the unique characteristics of ML workloads. 
On the one hand, \sysname leverages GPUDirect~\cite{gpudirect} to alleviate CPU overhead. 
On the other hand, \sysname employs control coalescing to make transport control decisions for every 32KB data chunk rather than every packet; this works well as many transport decisions such as CC do not require per-packet reaction, but just per-RTT to avoid overreaction~\cite{hpcc, falcon_slides}. 
As a result, \sysname can handle 400 Gbps unidirectional traffic with a single CPU core, and achieve the same message latency as hardware-based transport. 
\sysname leverages 256 RDMA QPs (Queue Pairs) per connection to perform multipathing, without worrying about the high QP context swapping overhead highlighted in previous work~\cite{srnic_nsdi23}; this is because ML workloads feature bulk data transfer with mostly MTU-sized large packets, which effectively amortizes the swapping overhead. 
\sysname's software transport is practical and economical in modern GPU servers, which usually have hundreds of powerful CPU cores; in fact, these CPU cores are often heavily underutilized. 
For example, from our private conversation with the model training team of a major GPU vendor, their CPU utilization is averagely 14.5\% out of 128 cores, when using an internal version of Megatron-LM~\cite{megatronlm} for model training. 

\sysname provides an expressive yet easy-to-use interface. To demonstrate the versatility of this interface and the power of \sysname's extensibility, we use three case studies. 
First, we implement a multipath transport protocol that mitigates flow collisions by leveraging packet spraying, i.e., randomly sending packets from a single connection across different paths~\cite{packet_spraying}.
Compared to existing RDMA transport on hardware, ML collectives using our transport achieve up to 4.5$\times$ higher throughput on NVIDIA ConnectX-7 NICs, and up to 1.9$\times$ higher on Broadcom Thor-2 NICs with rail-optimized topology. 
Second, we implement a receiver-driven protocol EQDS~\cite{eqds} to handle network incast in MoE-like workloads, achieving 4.9$\times$ better message tail latency over InfiniBand built-in transport. 
Third, we implement selective retransmission~\cite{selective_repeat} for efficient transport loss recovery, and show its superiority over RDMA hardware transport under packet loss. 
These case studies highlight that \sysname can effectively enable innovations in the transport protocols that are hard to implement in today's network stack without expensive and time-consuming changes. 
\sysname is open-sourced at \href{https://github.com/uccl-project/uccl}{https://github.com/uccl-project/uccl}. 
\presec
\section{Background and Motivation}
\postsec
\label{sec:background}

\subsection{GPU Networking: from RDMA to Collectives}
\postsub\label{ssec:gpu_net}
\vspace{0.05in}

GPU networking is extremely heterogeneous. 
At the high level, GPU collective communication libraries like NVIDIA NCCL~\cite{nccl} and AMD RCCL~\cite{rccl} use RDMA and kernel TCP (non-RDMA) for inter-server networking, with RDMA preferred since it is faster and more efficient. 
RDMA provides various communication primitives called Queue Pairs (QPs) including Reliable Connection (RC), Unreliable Connection (UC), and Unreliable Datagram (UD):
\begin{itemize}
    \item RC provides one-to-one message semantics (up to 1GB per operation), with NIC hardware handling packet reliability and CC. Some vendors like NVIDIA allow disabling CC. 
    \item UC also provides one-to-one message semantics but without NIC hardware logic for packet reliability or CC. 
    \item UD provides one-to-many datagram semantics, \ie, under one MTU per operation, without packet reliability or CC. 
\end{itemize}
Some cloud providers build their own RDMA NICs and QPs. For example, AWS EFA NICs replace RC with SRD (Scalable Reliable Datagram)~\cite{aws_srd} that implements datagram semantics, multipathing, packet reliability, and CC. 

To use RDMA NICs, the CPU issues \textit{verb} operations such as two-sided send/recv and one-sided read/write that transfer data over QPs. 
Internally, the verb builds a Work Queue Entry (WQE) and performs MMIO-write into the RDMA NIC registers. Upon completion, and depending on whether the verb is two-sided, the RDMA NIC generates a Completion Queue Entry (CQE) for the software to consume.
Note that UD supports only two-sided verbs, UC supports all verbs except RDMA read, while RC supports all verbs.
RDMA traffic could go through different network fabrics such as RoCE (RDMA over Converged Ethernet) and InfiniBand. 

\begin{figure}[t]
\centering
\prefig
\begin{minipage}[t!]{0.475\textwidth}
\includegraphics[width=\textwidth]{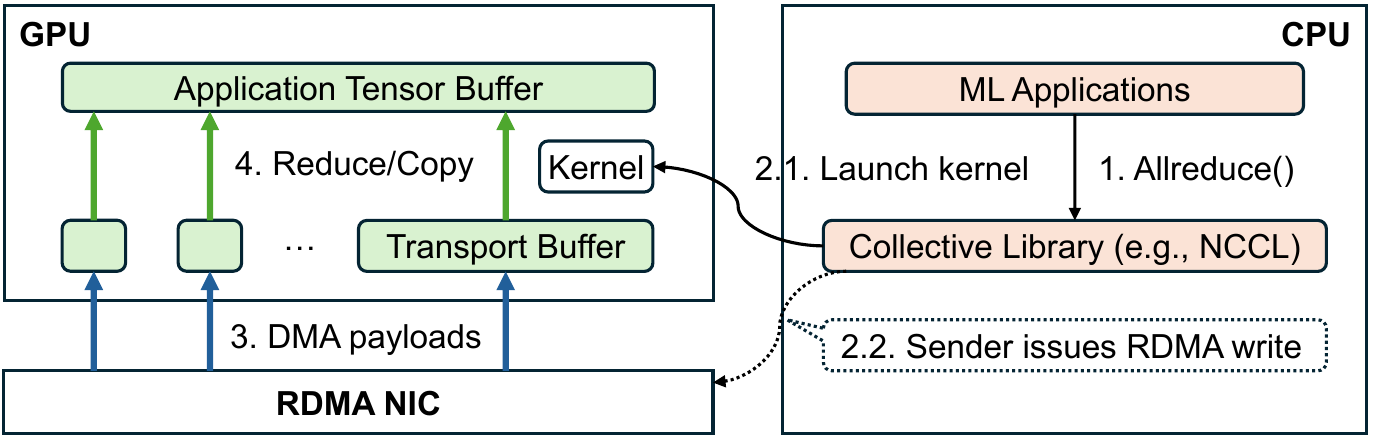}
\end{minipage}
\postfig
\caption{Collective communication over RDMA (receiver side). The receiver directly receives data into the GPU memory (\eg, GPUDirect~\cite{gpudirect}); the sender side works similarly except that it will additionally issue an RDMA write in step 2. }
\label{fig:gpu_net}
\postfigcaption
\vspace{0.05in}
\end{figure}

Figure \ref{fig:gpu_net} shows an overview of how collective communication uses RDMA. 
Once the ML application calls a collective like \allreduce, the collective library launches a \textit{reduction kernel} on each participant GPU to handle data reduction/copy. 
Next, the sender CPU issues multiple RDMA writes over RC QPs to transfer data chunk by chunk. The receiver CPU polls completion flags in its memory, which the sender sets upon write completions. 
The library manages a set of \textit{transport buffers} on the GPU memory to buffer the RDMA data, and relies on the GPU kernel to copy data between the transport buffers and application tensor buffers. 
The GPU kernel also performs reductions (\eg, sum, max) on the transport buffers (from multiple sender GPUs) into the tensor buffers. 

\presub
\subsection{Motivation on Extensibility}
\postsub\label{ssec:motivate_extend}

Host network transport on RDMA NICs is hard to evolve compared to software applications. This creates problems for fast-evolving ML workloads. 
We have shown two such examples in \S{\ref{sec:intro}};
below, we give four additional examples to motivate the need for transport layer extensibility. 

\para{Receiver-driven CC for incast.} Recent MoE serving workloads are prone to network incast problems. 
In DeepSeek's online deployment of their 671B V3 model~\cite{deepseek_v3}, each of the 320 GPUs holds a single expert module, where hidden states are exchanged between expert modules across GPUs, \ie, Expert Parallelism (EP). 
As the request pattern and load change over time, some experts become much hotter than others, receiving more network traffic from other experts leading to network incast issues. DeepSeek reports the hottest expert could receive 10$\times$ more load than the average one. 
Expert load balancing algorithms such as EPLB~\cite{eplb} try to balance the load by dynamically replicating experts.
However, this happens at a much slower pace (\eg, 10 minutes~\cite{deepep}) to avoid the high cost of moving experts, thus unable to handle transient incast. 
Such transient network incast can be better handled by receiver-driven CC~\cite{ndp, eqds}, which controls last-hop congestion---unfortunately, there is no receiver-driven CC on commercial RDMA NICs. 

\para{Application-transport codesigns.}
Codesigning applications and transport behaviors could bring huge performance benefits. 
For example, recent work MLT~\cite{mlt} customizes loss recovery behaviors for ML training to allow semi-reliable transmission based on the gradient importance from applications. 
Despite achieving great performance improvement, it is not feasible to integrate MLT into existing RDMA NICs even for the latest NVIDIA ConnectX-7~\cite{cx7}, due to a lack of enough programmability. 

\para{Inefficient loss recovery.} RDMA NICs are known to perform poorly under packet loss, especially for old-generation NICs~\cite{zeronic, irn, flor, srnic_nsdi23}. This is caused by the inefficient go-back-N retransmission logic hardcoded on these NICs due to limited on-chip SRAM constraints. As a result, RDMA deployment normally requires Priority Flow Control (PFC) to achieve a lossless network fabric. However, PFC may lead to deadlocks, head-of-line blocking, and victim flows~\cite{pfc_deadlocks, irn}, and its likelihood is higher as the GPU networking bandwidth keeps increasing (for the reason, we kindly refer to page 4 in~\cite{pfc_deadlocks}). 
If we could extend the transport layer of GPU networking with more efficient selective retransmission, we could better handle packet loss and rely less on PFC~\cite{irn}. 

\para{Heterogeneous NICs.} Datacenters usually consist of multiple generations and vendors of RDMA NICs due to continuous expansion, cost optimization, and to avoid vendor lock-in. While NVIDIA, Broadcom, AMD, and more vendors all have 400 Gbps RDMA NICs for ML~\cite{cx7, bcm_400g, amd_400g}, they come with subtly different control path logic such as packet reliability and CC. In practice, this heterogeneity reduces achievable bandwidth by 2-33$\times$ when communicating between NICs from different generations/vendors, as reported by Alibaba~\cite{flor}. 
Prior work Flor~\cite{flor} has shown that extensibly aligning these NICs' control path logic in software could avoid such a severe performance drop.

\presub
\subsection{Prior Work on Extensibility}
\postsub\label{ssec:bg_software}

\para{Leveraging SmartNICs.}
Several recent efforts aim to make RDMA transport programmable by offloading it to the SmartNIC RISC cores, but they have constraints in extensibility and performance. 
Google Falcon SmartNICs~\cite{falcon_slides, falcon_spec} only support programming rate update actions for latency-based Swift CC~\cite{swift} and path selection decisions with limited paths~\cite{plb}; similar constraints also apply to firmware updates on hardware RDMA NICs. 
AMD Pensando SmartNICs~\cite{amd_400g} support using P4 language~\cite{p4_ccr} to program their transport layer, but P4 has limited programmability, \eg, hard to implement efficient loss recovery; it is also unclear if they could support receiver-driven CC. 
FPGA-based SmartNICs provide higher performance but have limited extensibility due to hardware resource constraints~\cite{srnic_nsdi23}. 
AWS EFA SmartNICs~\cite{efa} implement a proprietary multipath reliable transport SRD~\cite{aws_srd} with out-of-order packet delivery using NIC ARM cores, to address network congestion in HPC and ML workloads. 
The SRD protocol is implemented using EFA-specific firmware and supports live upgrade with good extensibility. 
However, we empirically find that the EFA SmartNICs on AWS \texttt{p4d.24xlarge} GPU VMs perform poorly for connection-intensive \alltoall collectives (see \S\ref{ssec:collective_perf}). 
We attribute this to the limited processing power and cache capacity of SmartNIC ARM cores due to power constraints, as also demonstrated by prior research on publicly available SmartNICs~\cite{floem, ipipe, xenic_sosp21}. 

Note that the above \alltoall measurement uses the AWS EFA NICs on the \texttt{p4d.24xlarge} instances, and thus it may not apply to their new-generation EFA NICs on \texttt{p5/p5en/p6} instances. 
Specifically, the ever-changing EFA firmware and upgraded EFA hardware may lead to different performance results for connection-intensive \alltoall collectives. 
In general, if these SmartNICs get upgraded with higher processing power and cache capacity to better handle \alltoall, we view it as an echo to \sysname's high-level methodology of software extensibility for GPU networking. 

\begin{figure*}[t]
\centering
\prefig
\begin{minipage}[t!]{0.48\textwidth}
\includegraphics[width=\textwidth]{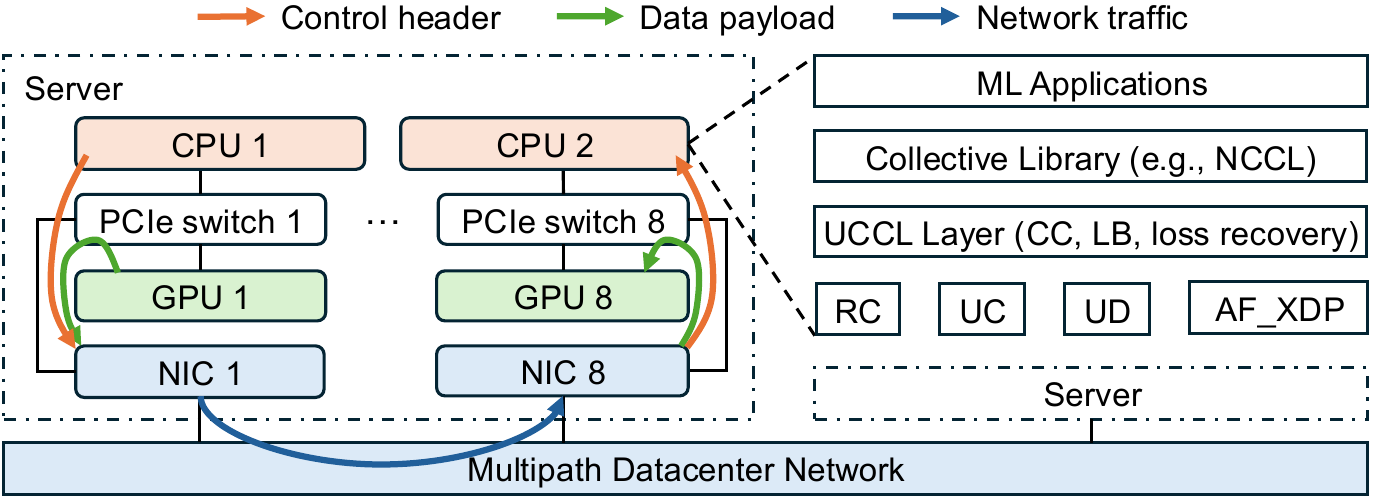}
\subcaption{\sysname architecture.}\label{fig:arch}
\end{minipage}
\hfill
\begin{minipage}[t!]{0.48\textwidth}
\includegraphics[width=\textwidth]{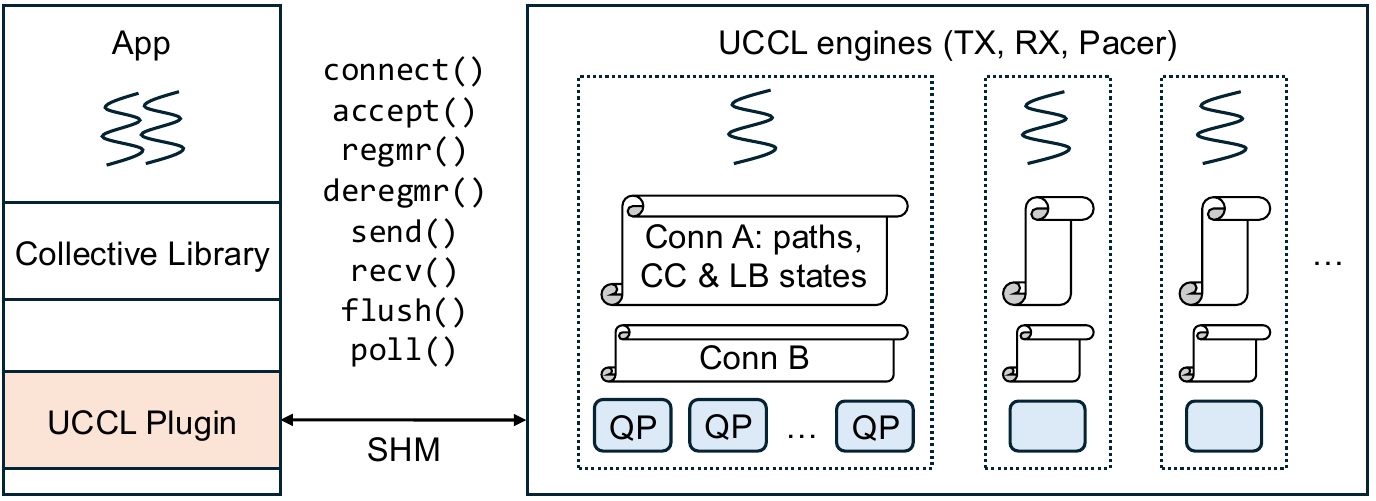}
\subcaption{\sysname threading model.}\label{fig:threading}
\end{minipage}
\postfig
\caption{Overview of \sysname extensible transport for GPU networking. We assume a common intra-server topology for GPU servers~\cite{dgx_box} where individual PCIe switches directly connect a few GPUs, NICs, and CPU, providing high-bandwidth data transfer among them.}
\postfigcaption\vspace{0.15in}
\end{figure*}


\para{Leveraging CPUs.}
A line of work has leveraged host CPUs to make better control decisions in GPU networking. 
ZeroNIC~\cite{zeronic} modifies the NIC hardware to run the control path of RDMA transport on CPUs, while leaving the data path on the NIC following GPUDirect. In contrast, \sysname aims to be more practical without modifying existing hardware; \sysname by design supports efficient multipathing, compared to ZeroNIC's single path transport. 
Flor~\cite{flor} leverages RDMA UC to bypass the control path of the RDMA hardware and implements flexible software control on CPUs. 
Flor targets CPU-based storage applications with 100 Gbps traffic per server, while \sysname targets more network-intensive ML applications that have 3.2+ Tbps traffic per server~\cite{aws_p5e, azure_h100} and develops multipathing to avoid flow collisions. 
To this end, \sysname adopts several different designs such as multi-QP and connection splitting (see \S\ref{ssec:multipath} and \S\ref{ssec:efficient_sw}). 

\para{Other efforts} design transport protocols to address specific network challenges. 
Ultra Ethernet Consortium (UEC)~\cite{uec} standardizes several multipath transport protocols with packet spraying to address flow collisions for ML workloads: a sender-driven one based on STrack~\cite{strack} and SMaRTT-REPS~\cite{smartt}, and a receiver-driven one based on EQDS~\cite{eqds}. 
Prior to UEC, MP-RDMA~\cite{mprdma} and MPTCP~\cite{mptcp} designed multipath protocols for CPU workloads to improve robustness under network failures. 
Overall, these protocols leverage various congestion signals such as Explicit Congestion Notification (ECN)~\cite{dctcp}, RTT~\cite{timely}, and packet trimming status~\cite{cp_nsdi14, ndp, eqds} to make multipath CC and LB decisions. 

\presec
\section{\sysname Design}
\postsec\label{sec:design}

Figure~\ref{fig:arch} shows the high-level architecture of \sysname. 
\sysname layer sits between the collective library such as NCCL and the low-level communication primitives exposed by NIC hardware, \eg, RC, UC, and UD for RDMA NICs and AF\_XDP~\cite{afxdp} (a user-space fast packet IO) for non-RDMA NICs. 
ML applications use collective APIs such as \texttt{\allreduce} and point-to-point APIs such as \texttt{SendRecv} exposed by the collective library without directly interacting with the \sysname layer. 
Both the collective library and the \sysname layer are compiled into individual shared libraries, \ie, \texttt{libnccl.so} and \texttt{libnccl-net.so} for NCCL, providing a drop-in replacement for ML applications without code modification or recompilation. 
\sysname leverages the network plugin system of existing collective libraries~\cite{nccl_plugin, rccl_plugin} to avoid changing the library code in most cases, with exceptions for \sysname over UD that requires a slight code modification (detailed in \S\ref{ssec:sw_hw_sep}). 
For brevity, the remaining paper targets RDMA NICs such as NVIDIA ConnectX NICs and AWS EFA NICs~\cite{efa}; we will explicitly mention when targeting non-RDMA NICs. 

Figure~\ref{fig:threading} shows the threading model of the \sysname layer. 
The \sysname plugin interacts with a group of \sysname engine threads via shared memory to create connections, register/deregister GPU memory regions with the RDMA NICs, and send/recv/flush/poll network messages.  
Each engine thread runs TX, RX, and pacing functionalities of \sysname multipath reliable transport for multiple \sysname connections. 
As described in \S\ref{ssec:sw_hw_sep}, \sysname engines instruct the RDMA NIC to receive network data, split control headers and application data payloads, and directly DMA them into the CPU and GPU memory separately. 
The whole process bypasses the packet reliability and CC logic on RDMA NIC hardware as much as possible by using proper RDMA primitives. 
After getting control headers in CPU, \sysname engines make transport decisions such as CC, LB, and handling packet loss and reordering. 
Since these decisions are executed by a normal user-space process on the CPU, instead of on RDMA NIC hardware, they can be easily extended by collective library or ML application developers.

In \sysname, all connections between a specific pair of NICs share the same set of QPs (\eg, 256), including the case of multiple GPUs sharing one NIC. This design fully harnesses the underlying multipath datacenter network while not burning too many QPs (\S\ref{ssec:multipath}). 
\sysname further integrates a bag of techniques to run software transport as efficiently as possible, \eg, control coalescing, connection splitting, chained posting, and more (\S\ref{ssec:efficient_sw}). 
\sysname also supports extending transport for non-RDMA NICs through the AF\_XDP user-space packet IO, bypassing conventional kernel TCP stacks. 
%
We now describe them in the rest of this section. 

\presub
\subsection{Separating Control Path and Data Path}
\postsub
\label{ssec:sw_hw_sep}

The overall goal of separating the control path and the data path is to enable running extensible transport on the CPU, while efficiently transferring data to/from GPUs in a GPUDirect manner. 
This goal has three specific aspects: 
(1) We should involve as little control logic as possible in the data path to let the CPU make more transport decisions, like CC and packet reliability. 
(2) We must achieve GPUDirect for the data path efficiency~\cite{ibm_vela, zeronic}.
(3) We should support heterogeneous RDMA NICs. For example, NVIDIA NICs support UC, while Broadcom and AWS EFA do not. 

\begin{figure}[t]
\centering
\prefig
\begin{minipage}[t!]{0.45\textwidth}
\includegraphics[width=\textwidth]{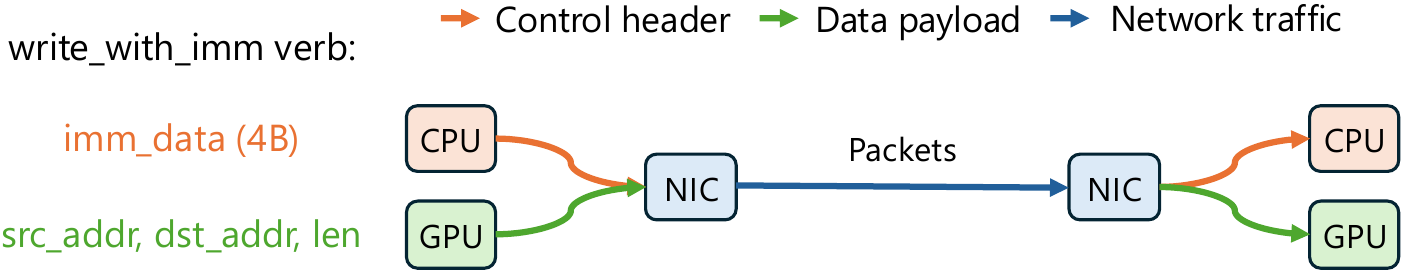}
\end{minipage}
\postfig
\caption{
Leveraging RDMA \texttt{write\_with\_imm} to separate control header and data payload for UC/RC. 
}
\label{fig:imm_uc}
\postfigcaption
\end{figure}

\para{UC as the preferred QP.}
\sysname chooses UC as the preferred QP whenever available on the RDMA NIC, similar to Flor~\cite{flor}, because it supports efficient segmentation and reassembly offloading to the NIC (in contrast to UD), while bypassing hardware-baked CC, loss recovery, and out-of-order packet handling (in contrast to RC). 
As shown in Figure~\ref{fig:imm_uc}, \sysname uses the efficient RDMA verb of write with immediate to transfer data chunks over UC; 
this verb operates in a two-sided mode, so that both sender and receiver CPUs can react to the data transfers. 
For the data path, the sender CPU specifies the addresses of source and destination data chunks when issuing the verb, and then the sender NIC will automatically segment the source data chunk into MTU-sized packets with packet headers prepended and send them out. 
When receiving these packets, the receiver NIC will remove the packet headers and reassemble payloads into the \textit{contiguous} memory region specified by the sender CPU. 

For the control path, write with immediate allows carrying a 32-bit \texttt{imm\_data} from the sender CPU to the receiver CPU, serving as the control header for \sysname transport. Figure~\ref{fig:imm_uc} shows an example.
UC guarantees that any successfully arrived chunk will generate a CQE with the \texttt{imm\_data} embedded, which is then consumed by the receiver CPU. 
Within the 32-bit budget, \sysname allocates 8 bits for connection ID and 7 bits for message ID, supporting 256 connections for each pair of NICs on a \sysname engine and 128 in-flight messages for each connection, which is sufficient for collective communication.
\sysname then allocates 8 bits to the chunk sequence number (CSN) to identify the position of a chunk in the message being transferred. 
Another 1 bit is allocated for marking the last chunk of a message. 
The remaining 8 bits are reserved for more advanced CC such as receiver-driven ones (see \S\ref{ssec:receiver_cc}). 

\para{RC with CC disabled.}
In practice, UC is not always supported across different RDMA NIC vendors~\cite{flor}, \eg, Broadcom~\cite{bcm_400g}. 
In these cases, \sysname will opt to RC with CC configured to be disabled, and then leverage RDMA write with immediate in a similar way as UC. 
On the one hand, RC prevents \sysname from customizing the packet reliability mechanism baked in NIC hardware;
on the other hand, it allows faster ACKs and more precise RTT estimation in hardware. 

\para{UD as the last resort.}
Some RDMA NICs do not allow disabling CC for their RC QPs, \eg, AWS EFA NICs (to be precise, EFA NICs do not have RC but only SRD). 
To support them, \sysname leverages UD at the cost of higher CPU usage compared to UC and RC. 
On the one hand, UD totally bypasses any hardware control logic on RDMA NICs, well-aligned with our goal. 
On the other hand, UD only supports sending/receiving MTU-sized data (\ie, no segmentation or reassembly offloading); so \sysname needs to consume more CPU cycles to do segmentation and reassembly. 

\para{\textit{Challenge of separation.}}
One key challenge for \sysname over UD is that UD does not support RDMA write with immediate (but only the \texttt{send/recv} verbs), so that \sysname over UD cannot designate the immediate data as the transport control header. 
Then \textit{how can \sysname separate control header and data payload using UD (\ie, placing them to CPU and GPU separately)}? 
\sysname must guarantee that the control header and data payload are \textit{fate-sharing} in terms of loss status and arrival order, so that \sysname can make valid transport decisions based on the control header. 
One strawman solution would be transferring the control header and data payload together as a single packet into the destination GPU memory, and then the CPU reading the control header from the GPU memory. But this incurs additional performance overhead. 

\begin{figure}[t]
\centering
\prefig
\begin{minipage}[t!]{0.48\textwidth}
\includegraphics[width=\textwidth]{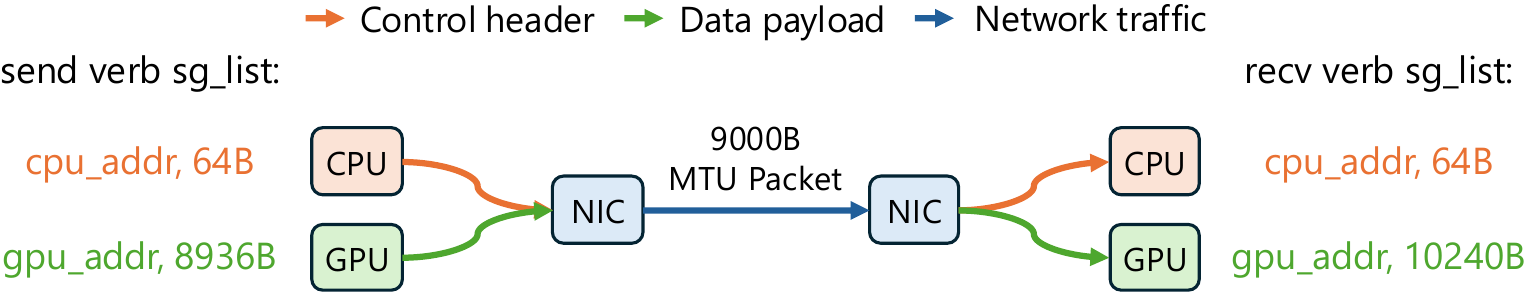}
\end{minipage}
\postfig
\caption{
Leveraging RDMA send/recv scatter-gather to separate control header and data payload for UD. 
}
\label{fig:sg_ud}
\postfigcaption
\end{figure}

\sysname's approach is to leverage the scatter-gather feature to let the NIC hardware automatically merge the control header and data payload during RDMA send, and split these two during RDMA recv; Figure~\ref{fig:sg_ud} shows an example. 
On the sender side, the CPU issues an RDMA send verb with a two-entry \texttt{sg\_list} that specifies the control header address+length on the CPU, and the data payload address+length on the GPU. 
The RDMA NIC will then read the header and payload from the CPU and GPU respectively, and merge them into a single network packet to send out, as long as the total length does not exceed the MTU size. 
On the receiver side, the CPU pre-posts a recv verb with a two-entry \texttt{sg\_list} that specifies the receiving address+length for the header and payload, respectively. 
Note that the header length must be a fixed value that the sender and receiver agree on, \eg, 64B in this example; the payload length specified in the recv verb need not exactly match the send verb, but should be no smaller. 
Later, when the packet arrives at the receiver NIC, the NIC will automatically split the header and payload across CPU and GPU, following the boundary of the fixed header length. 
\sysname's approach always keeps the control header fate-sharing with the data payload, and avoids the CPU reading any extra header from the GPU. 

\para{\textit{Challenge of reassembly.}}
\sysname over UD still faces another challenge of \textit{how to correctly and efficiently reassemble packets on the receiver GPU}. 
Recall that UD does not support reassembly offloading to the NIC, and only allows sending/receiving a single packet in one verb (\S\ref{ssec:gpu_net}). 
We note that sender-side segmentation is relatively easy, as the CPU could partition the transport sending buffer into individual data payloads (based on MTU size), and specify their addresses in send verbs. 
However, for the receiver-side reassembly, even if the CPU pre-posts recv verbs that specify in-order individual data payload addresses drawn from the transport receiving buffer, the packets will land into the buffer in an out-of-order manner due to packet loss or reordering over the multipath network. 
This reassembly challenge is unique to UD, as UC/RC allows the sender to directly specify the receiver-side GPU buffer addresses in the write with immediate verb. 

Addressing this challenge requires some form of \textit{scattered memcpy} GPU kernel that copies out-of-order data payloads to the transport receiving buffer (following the right order given by the receiver CPU). But the question is where to launch and run such a kernel. 
To avoid extra kernel launching overhead, \sysname chooses to fuse such scattered memcpy operations into the existing reduction kernel in collective libraries (\S\ref{ssec:gpu_net}). 
Our fused kernel will first do scattered memcpy to copy out-of-order data payloads into the transport buffers, then do the original reduction work from the transport buffers to the application tensor buffers. 
The only overhead of this approach is the extra GPU memory bandwidth consumption, but this is bounded by the network bandwidth. 
Given the high GPU memory bandwidth (\eg, 1.6-2.0 TB/s in A100), such extra bandwidth consumption is negligible. 

\para{For non-RDMA NICs},
\sysname builds a reliable transport atop UDP with the AF\_XDP technique, an efficient kernel socket that lets the NIC directly DMA network packets to user-space memory regions. 
We choose AF\_XDP as it achieves similar high performance as DPDK~\cite{dint}, but is kernel-native and does not require special NIC drivers, thus easy to deploy~\cite{ovs_dataplane}. 
Similar to how collective libraries such as NCCL use kernel TCP for non-RDMA NICs, \sysname over AF\_XDP does packet reassembly on the CPU, followed by a \texttt{cudaMemcpy()} to transfer the received message to the GPU. 

\presub
\subsection{Harnessing Multipath}
\postsub\label{ssec:multipath}

\begin{figure}[t]
\centering
\prefig
\begin{minipage}[t!]{0.475\textwidth}
\includegraphics[width=\textwidth]{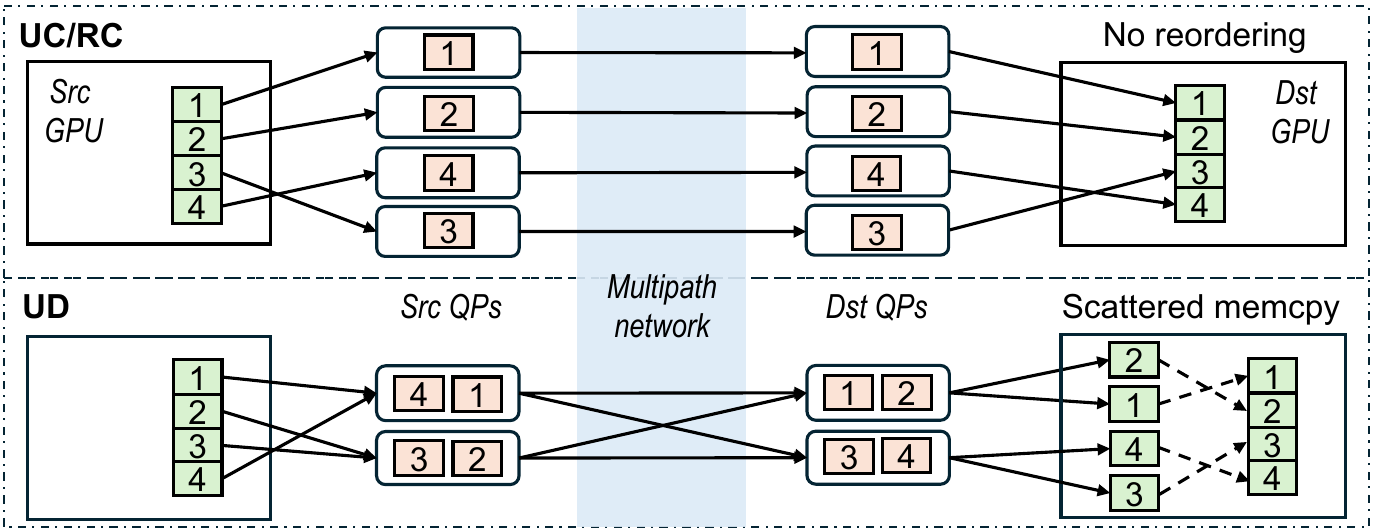}
\end{minipage}
\postfig
\caption{Multipathing and handling packet reordering in \sysname. }
\label{fig:reordering}
\postfigcaption
\end{figure}

\begin{figure}[!t]
\centering
\includegraphics[width=0.39\textwidth]
{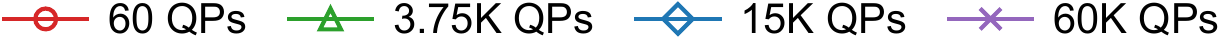}\vspace{0.05in}
\prefig
\begin{minipage}{0.48\textwidth}
    \begin{minipage}{0.48\textwidth}
        \centering
        \includegraphics[width=\linewidth]{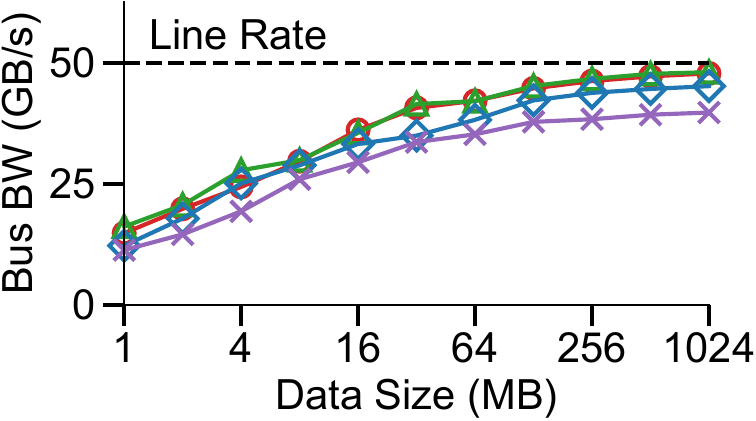}
        \subcaption{For RC QPs.}
        \label{fig:eval_qp_scale_rc}
    \end{minipage}
    \hfill
    \begin{minipage}{0.48\textwidth}
        \centering
        \includegraphics[width=\linewidth]{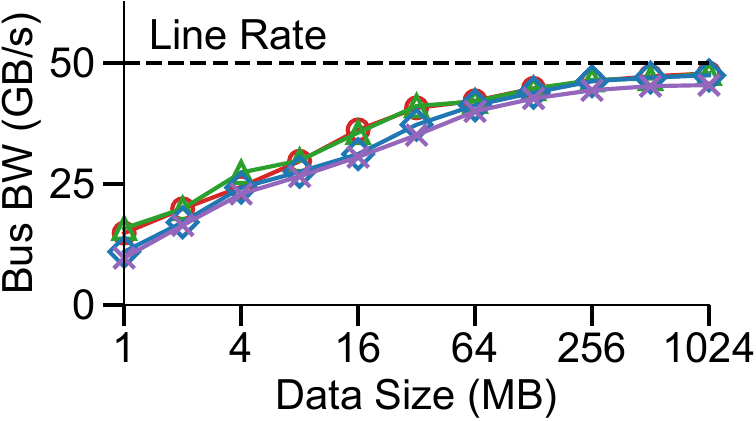}
        \subcaption{For UC QPs.}
        \label{fig:eval_qp_scale_uc}
    \end{minipage}
\end{minipage}
\postfig
\caption{\sysname \alltoall network bandwidth on \cx testbed (16 400G NICs, see \S\ref{sec:evaluation} for details) with different numbers of QPs per NIC. 
60 QPs per NIC correspond to the typical NCCL QP scaling factor of 4 (\ie, 4 QPs per connection) in production~\cite{llm_meta, strack}. 
60K QPs per NIC essentially model the QP swapping overhead for \sysname \alltoall across 241 GPUs using 256 paths per connection (\ie, 61440/256=240, plus 1 to include itself) or 961 GPUs using 64 paths. 
This has covered the largest-scale collective as far as we know, \ie, 256 GPUs by Meta~\cite{llm_meta} and 320 GPUs by DeepSeek-V3~\cite{deepseek_v3}. 
}
\label{fig:eval_vary_qp}
\postfigcaption
\end{figure}

One of the key motivations for GPU network extensibility is to harness the multipath capacity of modern datacenter networks (\S\ref{ssec:motivate_extend}). 
\sysname achieves this by using multiple UC, RC, or UD QPs, as shown in Figure~\ref{fig:reordering}. 
Basically, network traffic from different QPs will likely go through different network paths, as both RoCE and Infiniband usually use ECMP (Equal-Cost Multi-Path) for multipath routing with source and destination QP numbers as the hash inputs~\cite{llm_meta, ecmp_ib}. 
For UC and RC, \sysname by default uses 256 QPs, which provides maximum 256 different network paths as used by recent transport research~\cite{strack, smartt}. 
For UD, \sysname uses a much smaller number of QPs by combining different source and destination QPs. 
For example, 16 source UD QPs and 16 destination UD QPs will provide maximum 16$\times$16=256 different network paths, because for connection-less UD, each source QP can send packets to any destination QP. 
\sysname also supports a configurable number of QPs for different collectives, \eg, a smaller number might work well for \alltoall with relatively high entropy~\cite{llm_meta}. 
To avoid burning too many QPs, especially when the collective library creates multiple connections between the same pair of NICs (\eg, because of multiple GPUs sharing one NIC), \sysname lets all these connections \textit{share the same set of QPs}.

We note that making this multi-QP design choice is not trivial, especially as a line of prior work has highlighted the severe QP scalability issues on RDMA NICs~\cite{fasst_osdi16, erpc_nsdi19, scalerpc_eurosys19, flock_sosp21, collie_nsdi22, srnic_nsdi23, perf_atc24}. 
For example, SRNIC~\cite{srnic_nsdi23} reports $\sim$23\% bandwidth drop when scaling RC QPs from 256 to 512, and 46\% drop when scaling to 16k. 
This bandwidth drop is caused by the \textit{QP swapping overhead}: the NIC can only hold/cache limited QP contexts on its SRAM, and must spill/swap the excessive rest to the host DRAM over PCIe, incurring frequent QP swapping. 
Surprisingly, we do not observe such a severe performance drop for collective communication---Figure~\ref{fig:eval_vary_qp} shows only $\sim$17\% drop when scaling RC QPs from 60 to 60k, and a negligible drop for UC (that has a smaller QP context size than RC). 

There are two reasons behind this counter-intuitive phenomenon. 
First, ML workloads feature large messages, and thus collectives mostly transfer MTU-sized packets; such large transfers effectively amortize the QP swapping overhead. 
Second, with GPUDirect, GPU data transfers only go through the PCIe switch, but not the PCIe root complex that is connected to the CPU~\cite{manage_host_network}; therefore, there is no PCIe contention between GPU-NIC traffic and CPU-NIC traffic (that is incurred when NIC swaps QP contexts and CPU posts verbs). 
In contrast, prior work focuses on \textit{small-message CPU workloads} such as in-memory Key-Value stores that only transfer dozens of bytes at once for each QP, thus bottlenecked by the QP swapping overhead and PCIe traffic contention. 
Later in \S\ref{ssec:efficient_sw}, we show that \sysname optimizes transport efficiency by transferring data in a larger chunk granularity (\eg, 32KB), further reducing the QP swapping overhead.

For non-RDMA NIC multipathing, \sysname specifies different UDP ports in packets before sending them out in AF\_XDP. 
This adds no overhead compared to single-path transport. 

\para{Handling out-of-order packets.}
Many factors could cause out-of-order packet delivery, including multipathing, packet loss, and the unpredictable multi-QP scheduler in RDMA hardware~\cite{srnic_nsdi23}. 
Existing RDMA NICs perform poorly when handling out-of-order packets, as they cannot maintain large reordering buffers and states due to limited on-chip SRAM constraints~\cite{irn, srnic_nsdi23}. 
In contrast, \sysname is able to handle out-of-order packets efficiently thanks to its software flexibility and separation of data and control paths. 
Basically, \sysname follows typical TCP designs with \texttt{seq} and \texttt{ack} numbers to guide packet reordering, fast retransmission (upon duplicate ACKs), and timeout retransmission. 
\sysname sets a larger duplicate ACK threshold for fast retransmission, instead of the default three in TCP, to accommodate more frequent packet reordering caused by multipathing. 
Different from TCP, \sysname maintains its packet reordering buffers in the GPU memory and lets the NIC directly DMA network data there.
Figure~\ref{fig:reordering} depicts this process with examples. 
For UC/RC, the reordering buffers are individual data chunks, and the sender CPU specifies in-order chunk addresses when posting verbs. 
For UD, the reordering buffers are individual packet payloads, and the GPU reduction kernel reorders packets when copying them into the transport buffers (\S\ref{ssec:sw_hw_sep}). 

\presub
\subsection{Towards Efficient Software Transport}
\postsub
\label{ssec:efficient_sw}

So far, we have discussed how \sysname decouples the control path and data path to make flexible transport decisions on the CPU, and how \sysname achieves multipathing. 
The next question is \textit{how to efficiently implement a software multipath transport to support the high bandwidth in GPU networking}. 
This is challenging as a single GPU server could have 8$\times$400 Gbps RDMA NICs, totaling 3.2 Tbps bandwidth bidirectionally~\cite{azure_ec}; the next generation RDMA NIC will achieve 800 Gbps~\cite{nvidia_800g}, rendering 6.4 Tbps bandwidth. 
As a reference, Google's software transport Snap~\cite{snap_sosp19} can handle 80 Gbps traffic on a CPU core (though they do not use RDMA NICs). 
Our goal is to use 1 CPU core to handle 400G \textit{unidirectional} traffic (\ie, 2 cores for 400G bidirectional traffic; excluding possible pacer cores for receiver-driven CC). 
To this end, we leverage the following techniques: 

\para{Run-to-completion execution.}
Each \sysname engine thread runs RX, TX, pacing, timeout detection, and retransmission functionalities for a set of connections in an efficient run-to-completion manner~\cite{ix_osdi14, mos_nsdi}. 
\sysname employs Deficit Round Robin (DRR)~\cite{DRR} scheduling to fairly multiplex one engine thread among multiple functionalities and connections. 

\para{Connection splitting.}
To handle 400+ Gbps traffic per NIC more efficiently, \sysname pivots away from the Flor~\cite{flor} design of a single CPU core for one connection, but leverages multiple cores for one connection with connection splitting. 
Basically, \sysname equally partitions the 256 QPs among all engine threads responsible for a specific NIC; each engine thread gets its own connection states for CC and LB, forming a sub-connection. 
Within each sub-connection, \sysname uses RDMA SRQ and SCQ (Shared Recv/Completion Queues) to reduce the overhead when polling multiple recv and completion queues. 
The application threads atop the \sysname plugin are responsible for choosing the least-loaded engine (\eg, the engine with the least unconsumed messages) when dispatching messages via SHM. 
In this way, \sysname could scale transport processing of a single connection to multiple cores, and handle transient load imbalance among CPUs at runtime. 
It also reduces TX packet bursts by avoiding sending all messages at once from a single core. 

\para{Control coalescing.}
There is an inherent tradeoff between the control decision granularity and software transport efficiency. 
One could run CC, LB, and reliability logic for each packet to achieve precise control of the transport behaviors, at the cost of consuming more CPU cores. 
Alternatively, one could relax the control granularity by coalescing several same-path packets and making control decisions together, thus with lower CPU consumption.
For UC/RC, this also means an RDMA write could directly transmit several packets as a single data chunk, leveraging NIC-offloaded segmentation and reassembly. 
\sysname employs this control coalescing design with 32KB chunk size as default, striking a balanced tradeoff. 
Under this chunk size, \sysname can saturate 400 Gbps unidirectional bandwidth with 1 CPU core (\S\ref{sssec:cpu_scalability}), while not severely disrupting transport behaviors/performance (see our packet-level simulation in \S\ref{sssec:chunk_size}). 
Nevertheless, \sysname could also adaptively adjust chunk size based on the congestion level, \eg, switching to a small chunk size to make more precise control when congestion window (\texttt{cwnd}) drops below a threshold or severe packet loss happens. 

\para{Chained posting.}
UD does not support NIC offloading for segmentation and reassembly, thus it incurs more MMIO writes than UC/RC when issuing send/recv verbs (\eg, for individual packets). 
To reduce such overhead, \sysname leverages the chained posting feature of RDMA NICs to issue one MMIO write for posting up to 32 send/recv verbs. 
Concretely, the WQEs of these 32 verbs are chained together through the \texttt{next} pointer in previous WQEs, and get posted to the RDMA NIC in one MMIO write. 


\presub
\subsection{Congestion Signals}
\postsub\label{ssec:cc_signal}

Relatively restricted congestion signal is a common limitation for software-based reliable transport atop RDMA NICs, including \sysname and Flor~\cite{flor}. 
This is because existing RDMA NICs consume packet headers that contain congestion signals like ECN marks~\cite{dctcp} and packet trimming status~\cite{cp_nsdi14, ndp, eqds}, and deliver only the packet payload to the software. 
Fortunately, the software can still use the RTT congestion signal by leveraging hardware TX/RX timestamping supported by many RDMA NICs, and rely on packet loss as the last-resort congestion signal. 
Therefore, our current \sysname implementation uses per-path RTT and packet loss to detect congestion and choose paths. 
In fact, latency-based CC and LB are being widely used in Google's datacenters~\cite{swift, plb}. 

\para{Signal fidelity.} 
Running CC and LB in software using RTT also raises signal fidelity concerns. 
Overall, there are three factors affecting the fidelity: \circled{1} accurate RTT estimation at the sender, \circled{2} CC decision delay at either the sender or receiver, \ie, the software delay from receiving the congestion signal (\eg, ACK-derived timestamps) to updating the congestion window/rate, and \circled{3} ACK turnaround delay at the receiver, \ie, delay between receiving data chunks and sending back ACK. 
For \circled{1}, \sysname leverages the NIC hardware timestamps and excludes ACK turnaround delay from the RTT (similar to Swift~\cite{swift}). 
For \circled{2} and \circled{3}, theoretically, these delays impact how fast the sender can react to network condition changes, thus impacting the decision precision; however, in practice, even hardware-based transport handles CC events in a per-RTT granularity (\eg, tens of microseconds) rather than per-ACK to avoid overreaction~\cite{hpcc}. 
For example, Google Falcon hardware transport runs CC to update rate once per RTT~\cite{falcon_slides}. 
Thus, the few microseconds of decision delay introduced by the software is negligible. 
Nevertheless, \sysname still employs several techniques to reduce the two delays: similar to Flor~\cite{flor}, \sysname uses a dedicated high-priority QP for ACK (using in-network priority such as DSCP) and always first polls its completion queue; \sysname further allocates the ACK polling a higher processing budget during DRR scheduling. 
We quantify these delays in \S\ref{sssec:eval_cc_signal}. 

\presec
\section{Extensibility Case Studies}
\postsec\label{sec:case_study}

\sysname provides expressive interfaces to implement and extend multipath transport. 
Due to space limitations, we elaborate on them in Appendix~\ref{ssec:interface}. 
Collective library and application developers could also directly extend \sysname transport code, \eg, with the new loss recovery scheme in MLT~\cite{mlt}, and deploy it quickly in a normal user-space process. 
We now demonstrate how \sysname extensibility enables new transport designs that work best for different ML workloads. 

\presub
\subsection{Multipath Transport with Packet Spraying}
\postsub\label{ssec:case_multipath}

Recent transport research~\cite{strack, smartt} and UEC advocate packet spraying with hundreds of paths as an effective way to address flow collisions in ML workloads. 
It is naturally challenging for hardware NICs to implement packet spraying because of excessive per-path states, such as path RTTs. 
Instead, \sysname can easily support packet spraying by maintaining per-path RTT in software. 
\sysname's software transport uses Power-of-Two sampling~\cite{power_of_two} to select a path with the lowest RTT, then runs CC to decide how many packets and what rate to transmit. 
\sysname implements two CC algorithms: one is CUBIC~\cite{tcp_cubic} used in Linux kernel TCP as the default CC, and another is Swift~\cite{swift}, an RTT-based CC used by Google. 
\sysname supports both per-path CC states (\eg, per-path \texttt{cwnd}) and global CC states controlling all paths; both achieve similar collective performance in our testbeds. \sysname by default uses global CC during evaluation. 

\presub
\subsection{Receiver-Driven CC}
\postsub\label{ssec:receiver_cc}

Receiver-driven transports, such as EQDS~\cite{eqds}, NDP~\cite{ndp}, and Homa~\cite{homasigcomm}, proactively control packet sending rate at the receiver by allocating credits to senders. These transports have proven to resolve the last-hop congestion for network incast effectively, which could happen in MoE serving (\S\ref{ssec:motivate_extend}). 
However, to the best of our knowledge, there are no off-the-shelf NICs that support receiver-driven transports. 
One reason is that they are vastly different from popular sender-driven ones, and implementing them would require NIC hardware modification. 
Instead, \sysname's extensibility enables developers to quickly implement and tune receiver-driven transports in software. 
We choose to implement EQDS~\cite{eqds} in \sysname as an example, the state-of-the-art receiver-driven transport adopted by UEC~\cite{uec}. 
Our EQDS implementation closely follows the EQDS paper~\cite{eqds} with a dedicated pacer thread per NIC on the receiver to issue credit packets for senders. For more details, we refer to Appendix~\ref{sec:eqds_impl}.

\presub
\subsection{Efficient Loss Recovery}
\postsub\label{ssec:loss_recovery}

\sysname allows customizing the transport loss recovery logic to support more advanced mechanisms other than the go-back-N retransmission baked into many RDMA NICs. 
Go-back-N directly drops out-of-order packets to avoid buffering them on expensive on-chip SRAM, but this gives poor performance when packet loss happens~\cite{erpc_nsdi19, srnic_nsdi23, dcqcn}. 
Instead, we implement a more efficient selective retransmission~\cite{selective_repeat} in \sysname, by maintaining reordering buffers in the GPU memory (\S\ref{ssec:multipath}).
Our implementation follows the standard selective retransmission mechanism in TCP, and uses a \texttt{std::map} to track an arbitrary number of out-of-order packets. 
With more efficient loss recovery in \sysname, ML workloads could possibly run in lossy datacenter networks without PFC (\S\ref{ssec:motivate_extend}). 

\begin{table*}[t]
\prefig
\begin{center}
{\footnotesize
\begin{tabular}{cccccccc}
    \toprule
    \makecell{Name}
    &\makecell{\# of Servers}
    &\makecell{Network}
    &\makecell{Topology}
    &\makecell{MTU}
    &\makecell{NIC}
    &\makecell{GPU}
    &\makecell{CPU}
    \\\toprule
    CX\_IB & 2 & InfiniBand & Same rack & 4KB & NVIDIA ConnectX-7 400G $\times$8 & NVIDIA H100-80G $\times$8 & 128 cores \\
    CX\_ETH & 6 & Ethernet & Cross racks, fat-tree & 4KB & NVIDIA ConnectX-7 400G $\times$8 & NVIDIA H100-80G $\times$8 & 160 cores \\
    AMD & 4 & Ethernet & Cross racks, rail-optimized & 4KB  & Broadcom Thor‑2 400G $\times$7 & AMD MI300X-192G $\times$7 & 128 cores \\
    EFA & 4 & Ethernet & Cross racks, fat-tree & 9KB & AWS EFA 100G $\times$4 & NVIDIA A100-40G $\times$8 & 96 cores \\
    \toprule
\end{tabular}
}
\end{center}
\caption{Evaluation testbeds. EFA is rented from AWS with \texttt{p4d.24xlarge} instances. For brevity, we also refer to ConnectX‑7 as CX‑7. 
}
\postfigcaption\vspace{-0.14in}
\label{tab:eval_testbed}
\end{table*}

We note that the latest generation of NVIDIA RDMA NICs~\cite{cx7} has implemented a limited form of selective retransmission with a small fixed tracking window (to keep low SRAM usage). 
It will fall back to go-back-N when the number of unacked inflight packets goes beyond that window; in ML clusters with increasing networking bandwidth and thus more inflight packets, such hardware-based loss recovery can hardly work as effectively as the flexible software-based one in \sysname. 

\presec
\section{Implementation} 
\label{sec:implementation}
\postsec

We implement \sysname in 28.4K lines of C++ as a network plugin to NCCL/RCCL, the standard collective library for NVIDIA/AMD GPUs. 
Our current implementation supports NVIDIA RDMA NICs via RC and UC, Broadcom RDMA NICs via RC, AWS EFA NICs via UD/SRD, and AWS ENA non-RDMA NICs via AF\_XDP. 
\sysname leverages the standard \texttt{libibverbs} for RDMA NICs and kernel built-in AF\_XDP for non-RDMA NICs; therefore, it should naturally support other vendors' NICs. 
To enable scattered memcpy when using UD for AWS EFA NICs, we add or modify $\sim$170 LoC to the NCCL codebase, including a scattered memcpy GPU kernel and C++ code that passes packet pointers to the kernel via CPU-GPU shared memory (created by \texttt{cudaHostAlloc()}). 
We added two new interfaces to the NCCL network plugin system: \texttt{irecv\_scattered} that receives data in a list of scattered packets, and \texttt{irecv\_free\_ptrs} that frees the scattered packet buffers after scattered memcpy. \sysname also supports running NCCL with multiple processes. 

There are some implementation details worthy of note. 
For NVIDIA NICs, to calculate network RTT based on NIC hardware TX/RX timestamps and CPU ACK timestamps, we implement a NIC-CPU clock synchronization scheme following Swift~\cite{swift}. We use a synchronization time interval of 100$\mu s$. 
For AWS EFA NICs that do not support hardware timestamps yet, we calculate RTT based on software timestamps minus estimated packet queueing delay at the host. We estimate the queueing delay by dividing the size of queued packets by the NIC bandwidth. 
For AWS EFA NICs, we find that allocating UD QPs with consecutive QP numbers for a connection yields much better performance than not. We suspect this is because the EFA NICs do round-robin assignments when mapping a QP (and its associated packets) to a ARM core; consecutive allocations help map the QPs of a connection to different ARM cores, avoiding load imbalance. 

\presec
\section{Evaluation}
\postsec
\label{sec:evaluation}

In this section, we aim to answer the following questions: 
\begin{itemize}
    \item What is the collective performance of \sysname software multipath transport compared to hardware-based ones (\S\ref{ssec:collective_perf})?
    \item Can \sysname improve ML application performance (\S\ref{ssec:app_perf})?
    \item Can \sysname extensibility benefit certain workloads (\S\ref{ssec:extensibility_eva})?
    \item What is the scalability of \sysname (\S\ref{ssec:scalability})? 
    \item How do different designs impact \sysname (\S\ref{ssec:design_drill})?
\end{itemize}

\para{Testbed.}
Table \ref{tab:eval_testbed} describes our evaluation testbeds. 
The \cx testbed under the same rack mainly serves as a performance stress-testing for \sysname software implementation, as it has no network congestion or flow collisions. 
The \cxeth testbed across two racks evaluates how \sysname handles network congestion and flow collisions. 
The AMD and EFA testbed across racks evaluate the genericity of \sysname (\ie, applying to AMD GPUs and Broadcom/EFA NICs). 
Each AMD server has one NIC down out of 8; therefore, we only use 7 GPUs for evaluation. 
For some tests, we try to simulate a larger testbed by disabling NCCL NVLink (or RCCL xGMI) and SHM communication. 
This forces GPUs within a server to communicate through the network, and each GPU essentially behaves like a virtual server, \eg, \cx becomes 16 virtual GPU servers each with 400G bandwidth.
%

\para{Experiment setup.}
For collective performance evaluation, we use NCCL v2.23.4~\cite{nccl_v2_23_4} and NCCL-tests \texttt{9d26b84}~\cite{nccl_tests}, the latest versions when we start integrating \sysname into NCCL, and focus on the representative \allreduce and \alltoall collectives. 
NCCL-tests vary the collective message size and measure the achieved bus bandwidth. We focus on message sizes from 1MB to 1GB, commonly used in real ML workloads~\cite{llm_alibaba, simai, autoccl}. 
For AMD GPUs, we use RCCL \texttt{532f54c}~\cite{rccl_code} and RCCL-test \texttt{5b27b96}~\cite{rccl_tests}. 

\para{Comparison baselines.}
On the \cx testbed, we compare \sysname to the NCCL built-in RDMA support that uses CX-7 RC. 
\sysname uses 256 UC/RC QPs for each NIC pair, while the NCCL built-in uses QP scaling with 4 RC QPs per connection~\cite{llm_meta, strack}. 
\sysname uses CUBIC CC on \cx unless specified, which performs better than Swift in the lossless InfiniBand. 
Since InfiniBand has no packet loss for CUBIC to reduce \texttt{cwnd}, to avoid severe network congestion, we enforce a maximum \texttt{cwnd} value in \sysname CUBIC CC to limit the maximum inflight bytes, similar to TCP flow control. 

On the EFA testbed, we compare \sysname to the official AWS NCCL-EFA plugin v1.13.2~\cite{aws_nccl_plugin} that uses the multipath SRD~\cite{aws_srd}. 
\sysname uses 10$\times$26=260 UD QPs to serve all connections, which yields the best performance empirically, while the NCCL-EFA plugin uses the best parameters recommended by AWS. 
\sysname uses CUBIC CC on EFA by default, as EFA NICs do not support hardware timestamping that is critical to Swift CC. 
In both testbeds, \sysname uses per-path RTT for multipath LB (\S\ref{ssec:case_multipath}). 

As the cost of extensible software transport, NCCL atop \sysname uses more CPU cores than vanilla NCCL. 
By default, the vanilla NCCL uses 2 cores per GPU, while \sysname uses 2 more cores per NIC to run the engine threads, with 1 additional core per NIC for receiver-driven CC due to the pacer. 

\presub
\subsection{Collective Performance}
\postsub\label{ssec:collective_perf}

\begin{figure}[!t]
\centering
\includegraphics[width=0.32\textwidth]
{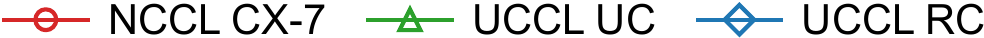}\vspace{0.05in}
\prefig
\begin{minipage}{0.48\textwidth}
    \begin{minipage}{0.48\textwidth}
        \centering
        \includegraphics[width=\linewidth]{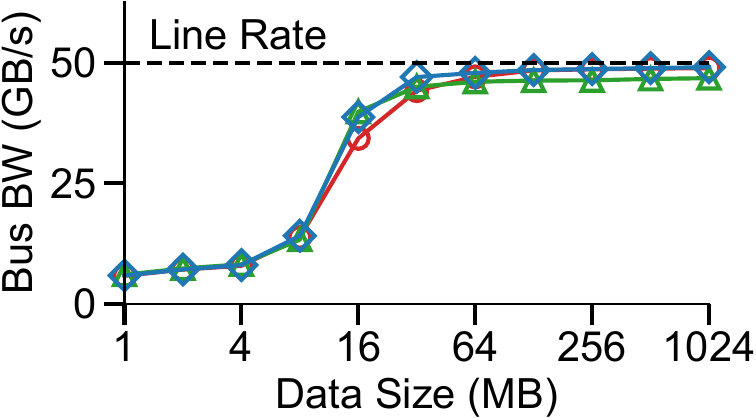}
        \postfig\vspace{-0.1in}
        \subcaption{\uallreduce.}
        \label{fig:eval_cxib_allreduce}
    \end{minipage}
    \hfill
    \begin{minipage}{0.48\textwidth}
        \centering
        \includegraphics[width=\linewidth]{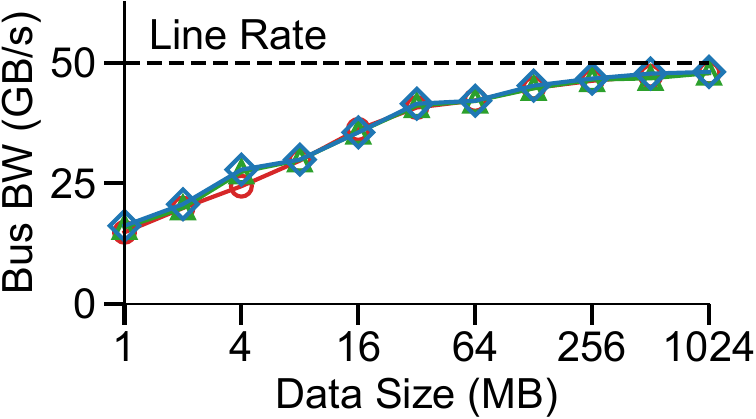}
        \postfig\vspace{-0.1in}
        \subcaption{\ualltoall.}
        \label{fig:eval_cxib_alltoall}
    \end{minipage}
\end{minipage}
\vspace{-0.10in}
\caption{NCCL-tests results on \cx (NVLink+SHM disabled to simulate a larger testbed).}
\label{fig:eval_cxib_collectives}
\postfigcaption
\end{figure}

\para{On \cx.}
Figure~\ref{fig:eval_cxib_collectives} compares the \allreduce and \alltoall performance of \sysname vs. ConnextX-7 on the \cx testbed. 
\sysname UC/RC performs almost the same as ConnectX-7 with various data sizes, except that \sysname UC performs slightly worse (\ie, less than 4\%) than ConnextX-7 and \sysname RC for \allreduce after 128MB.
For \allreduce, this exception is because of the extra overhead of handling packet reliability in UC; 
for \alltoall, ConnextX-7 and \sysname RC do not perform better than \sysname UC, because \alltoall is connection-intensive, causing more QP swapping for RC. 
These experiments confirm that \sysname can make highly efficient control decisions in software, reaching the performance of ASIC-based RDMA NICs (\ie, NVIDIA ConnectX-7) for ML collectives. 

\begin{figure}[!t]
\centering
\includegraphics[width=0.42\textwidth]
{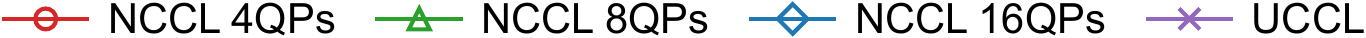}\vspace{0.05in}
\prefig
\begin{minipage}{0.48\textwidth}
    \begin{minipage}{0.48\textwidth}
        \centering
        \includegraphics[width=\linewidth]{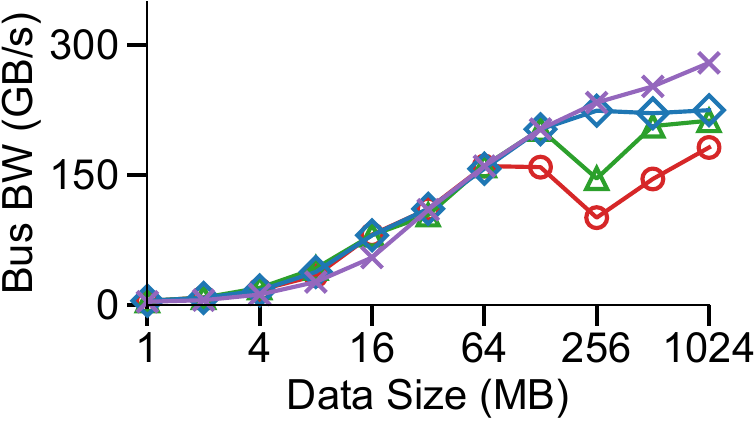}
        \postfig\vspace{-0.1in}
        \subcaption{\uallreduce.}
        \label{fig:eval_cxeth_allreduce}
    \end{minipage}
    \hfill
    \begin{minipage}{0.48\textwidth}
        \centering
        \includegraphics[width=\linewidth]{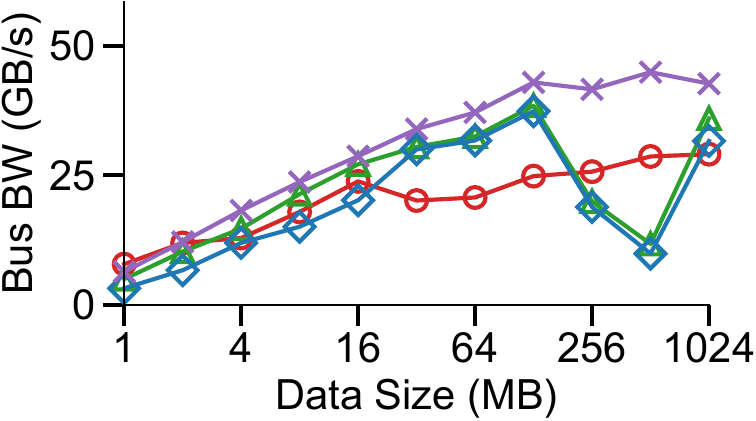}
        \postfig\vspace{-0.1in}
        \subcaption{\ualltoall.}
        \label{fig:eval_cxeth_alltoall}
    \end{minipage}
\end{minipage}
\vspace{-0.1in}
\caption{NCCL-tests results on \cxeth (NVLink+SHM enabled). }
\label{fig:eval_cxeth_collectives}
\postfigcaption
\end{figure}

\para{On \cxeth.}
Figure~\ref{fig:eval_cxeth_collectives} compares the performance of \sysname vs. ConnextX-7 on the \cxeth testbed. 
For NCCL over CX-7, we use multiple QPs for each connection between two GPUs by adjusting \texttt{NCCL\_IB\_QPS\_PER\_CONNECTION}. 
However, it still suffers from significant performance drops under larger message sizes and lower peak performance. 
This is because flow collisions in the fat-tree topology cause severe network congestion, which in turn leads to exponential backoffs in the CX-7 CC mechanism. 
Instead, \sysname is able to scale up performance stably with message size increasing, by making smarter LB and CC decisions in software. 
Overall, \sysname outperforms CX-7 by up to 2.32/1.60/1.24$\times$ (corresponding to 4/8/16 QPs) and 1.79/3.82/4.54$\times$ for \allreduce and \alltoall, respectively.

\begin{figure}[!t]
\centering
\includegraphics[width=0.4\textwidth]
{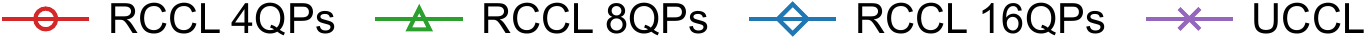}\vspace{0.05in}
\prefig
\begin{minipage}{0.48\textwidth}
    \begin{minipage}{0.48\textwidth}
        \centering
        \includegraphics[width=\linewidth]{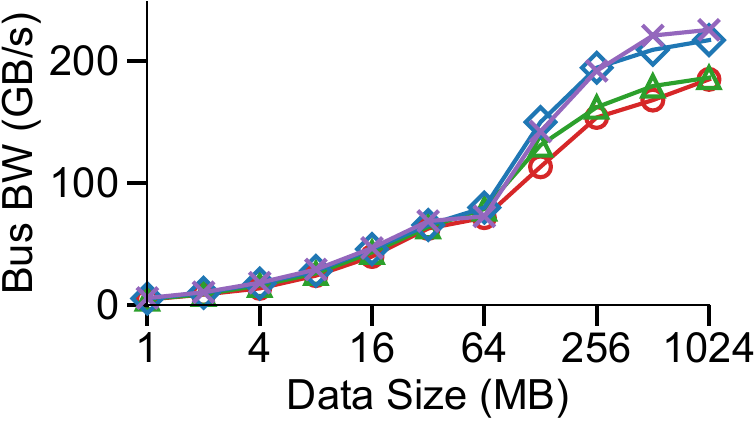}
        \postfig\vspace{-0.1in}
        \subcaption{\uallreduce.}
        \label{fig:eval_amd_allreduce_pxnon}
    \end{minipage}
    \hfill
    \begin{minipage}{0.48\textwidth}
        \centering
        \includegraphics[width=\linewidth]{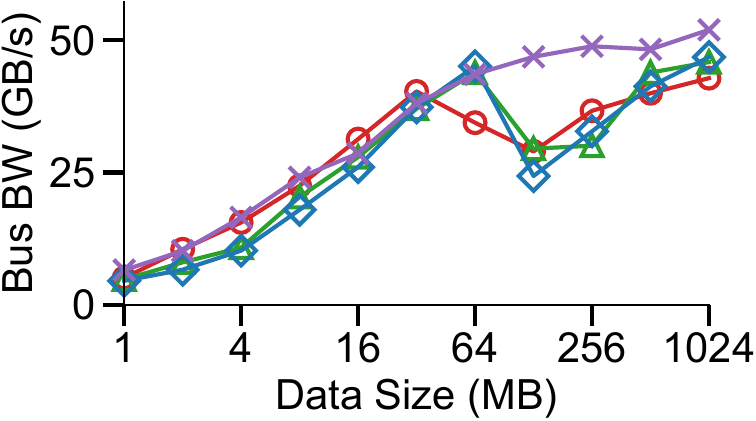}
        \postfig\vspace{-0.1in}
        \subcaption{\ualltoall.}
        \label{fig:eval_amd_alltoall_pxnon}
    \end{minipage}
\end{minipage}
\vspace{-0.1in}
\caption{RCCL-tests results on AMD (xGMI+SHM enabled).}
\label{fig:eval_amd_collectives_pxnon}
\postfigcaption
\vspace{0.05in}
\end{figure}

\para{On AMD.}
Figure~\ref{fig:eval_amd_collectives_pxnon} shows the performance comparison on the AMD testbed. 
This testbed has a rail-optimized topology that helps reduce network congestion when used together with the PXN technique~\cite{pxn}. 
Even under this topology, \sysname still outperforms Thor-2 by up to 1.34/1.23/1.08$\times$ (corresponding to 4/8/16 QPs) and 1.62/1.62/1.93$\times$ for \allreduce and \alltoall, by better controlling network congestion. 

\begin{figure}[!t]
\centering
\includegraphics[width=0.45\textwidth]
{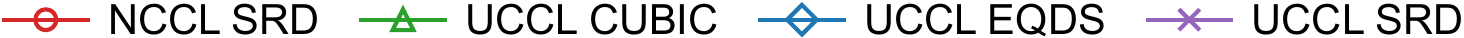}
\vspace{0.05in}
\prefig
\hfill
\begin{minipage}{0.48\textwidth}
    \begin{minipage}{0.48\textwidth}
        \centering
        \includegraphics[width=\linewidth]{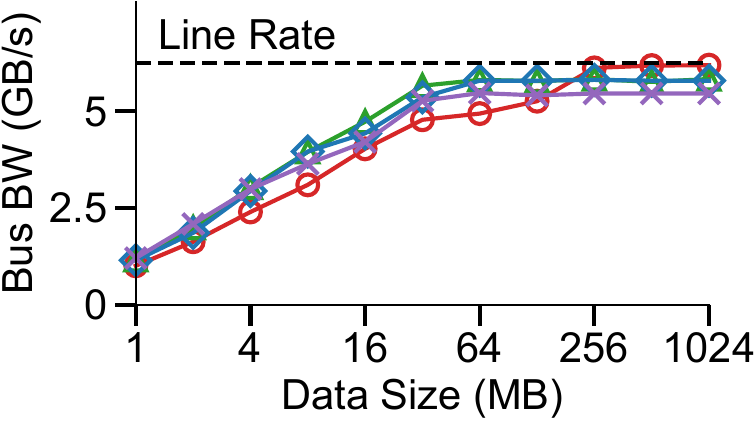}
        \postfig\vspace{-0.1in}
    \subcaption{\uallreduce.}\label{fig:eval_efa_allreduce}
    \end{minipage}
    \hfill
    \begin{minipage}{0.48\textwidth}
        \centering
        \includegraphics[width=\linewidth]{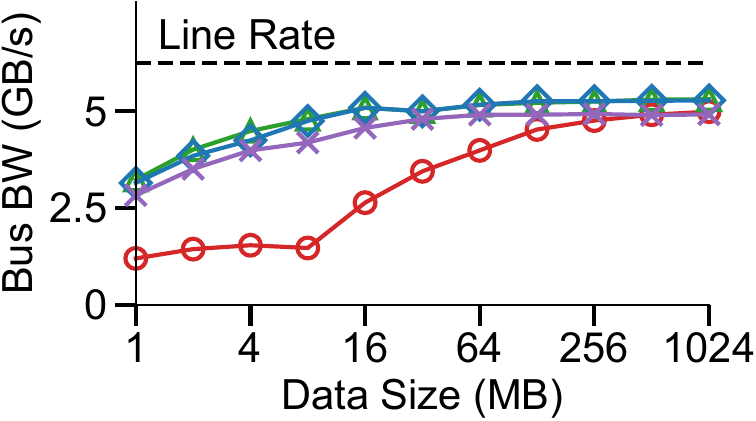}
        \postfig\vspace{-0.1in}
    \subcaption{\ualltoall.}\label{fig:eval_efa_alltoall}
    \end{minipage}
\end{minipage}
\vspace{-0.10in}
\caption{NCCL-tests results on EFA (NVLink+SHM disabled).}
\label{fig:eval_efa_collectives}
\postfigcaption\vspace{0.05in}
\end{figure}

\para{On EFA.}
Figure~\ref{fig:eval_efa_collectives} compares the collective performance of \sysname and SRD on the EFA testbed. 
\sysname CUBIC and EQDS achieve similar performance with significant improvement over SRD (except for $\ge$256MB \allreduce): for \allreduce, \sysname outperforms SRD by up to 1.27$\times$; for \alltoall, the speedup is up to 3.27$\times$. 
\sysname outperforms SRD because beefy CPU cores are faster in making transport decisions than wimpy ARM cores on \texttt{p4d.24xlarge} EFA NICs (that run SRD), especially when handling connection-intensive \alltoall. 
Surprisingly, simple CUBIC CC performs very well. This is mainly because \sysname leverages hundreds of network paths and avoids congested ones before sending packets; so most of the time, CC does not get involved. 
The small performance gap after 256MB is caused by the extra control header added by \sysname, while SRD directly reuses the UDP header as its control header (\sysname cannot reuse as it is not exposed to CPU software). 

\sysname could also directly leverage the SRD protocol of EFA NICs to send/recv individual datagram packets, denoted as \sysname SRD. 
\sysname SRD does connection management and multipath load balancing in CPU, and packet reassembly in GPU (\S\ref{ssec:sw_hw_sep}), overcoming the limited processing power and cache capacity of SmartNICs on \texttt{p4d.24xlarge}. 
As a result, it achieves similar high performance to \sysname CUBIC and EQDS. 

\begin{figure}[!t]
\centering
\includegraphics[width=0.24\textwidth]
{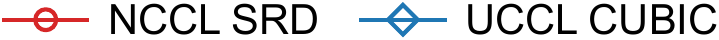}\vspace{0.05in}
\prefig
\begin{minipage}{0.48\textwidth}
    \begin{minipage}{0.48\textwidth}
        \centering
        \includegraphics[width=\linewidth]{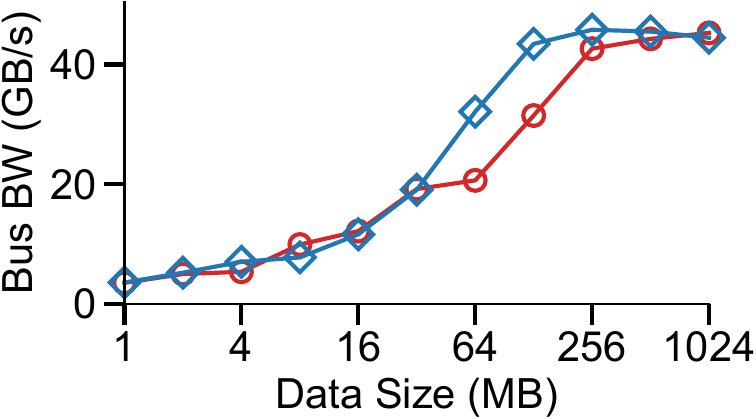}
        \postfig\vspace{-0.1in}
        \subcaption{\uallreduce. }
        \label{fig:eval_efa_nvlink_allreduce}
    \end{minipage}
    \hfill
    \begin{minipage}{0.48\textwidth}
        \centering
        \includegraphics[width=\linewidth]{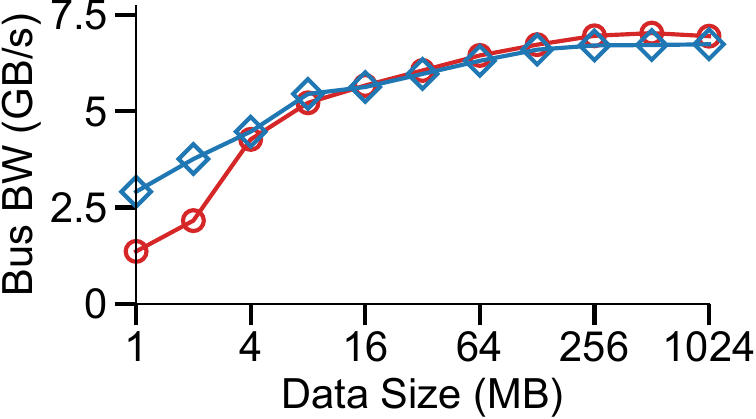}
        \postfig\vspace{-0.1in}
        \subcaption{\ualltoall.}
        \label{fig:eval_efa_nvlink_alltoall}
    \end{minipage}
\end{minipage}
\vspace{-0.10in}
\caption{NCCL-tests results on EFA (NVLink+SHM enabled).}
\label{fig:eval_efa_nvlink_collectives}
\postfigcaption
\vspace{0.1in}
\end{figure}

Figure~\ref{fig:eval_efa_nvlink_collectives} shows the collective performance comparison on the EFA testbed when NVLink and SHM are enabled (meaning a smaller testbed). 
Even though a major amount of data traffic goes through the high-bandwidth NVLink, \sysname still achieves much higher or comparable collective performance than SRD, \ie, up to 1.57$\times$ for \allreduce and 2.14$\times$ for \alltoall. 

Due to space limitations, we show the results of more collectives in Appendix~\ref{ssec:more_coll} and the results on non-RDMA NICs in Appendix~\ref{ssec:afxdp_perf} (up to 4.1$\times$ higher performance). 

\presub
\subsection{Application Performance}
\postsub\label{ssec:app_perf}

\sysname is motivated by the practical needs of both LLM training and serving workloads. 
We run two applications to evaluate how \sysname improves ML workload performance on the EFA testbed. 
One is a classic ResNet~\cite{resnet} distributed training in PyTorch with the widely-used 1F1B mechanism~\cite{pipedream2028, poseidon_atc} that tries to overlap communication with computation. However, as the GPU compute capability grows much faster than the network bandwidth, the communication time cannot be entirely hidden by the computation time, especially on EFA. 
Another is a DeepSeek-V3-like MoE serving application. 
However, directly serving DeepSeek-V3 on our testbed is impractical, because (1) it uses a customized communication library rather than the standard NCCL, so it cannot directly use \sysname, and (2) it requires a large number of GPUs to run meaningful expert parallelism, \eg, 320 GPUs by DeepSeek~\cite{deepseek_v3} and 96 GPUs by SGLang~\cite{sglang_ep}. 
Instead, we get a realistic DeepSeek-V3 trace including GPU compute time and network message sizes (\ie, hidden state sizes) from~\cite{deepseek_sim}, and emulate the computation and communication behaviors with PyTorch and NCCL. 

\begin{figure}[!t]
\centering

\prefig
\begin{minipage}{0.23\textwidth}
    \centering
    \includegraphics[width=\linewidth]{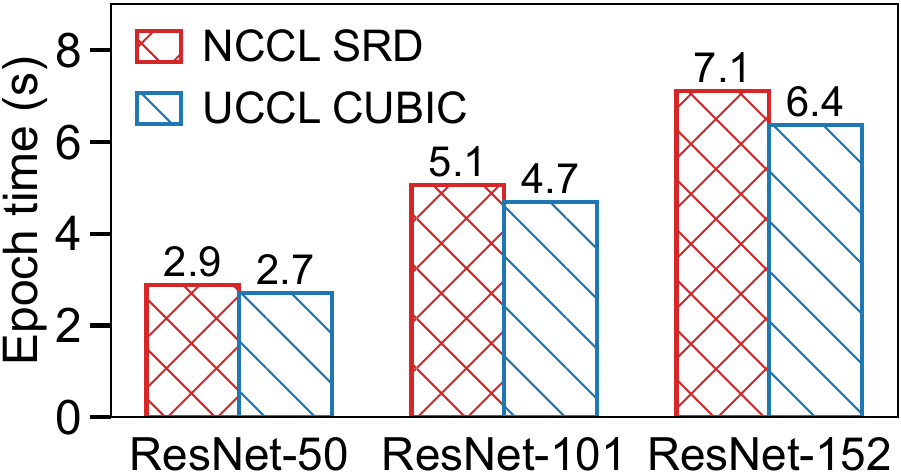}
    \subcaption{ResNet distributed training.}
    \vspace{0.18in}
    \label{fig:eval_app_resnet}
\end{minipage}
\hfill
\begin{minipage}{0.23\textwidth}
    \begin{minipage}{\textwidth}
        \centering
        \includegraphics[width=\linewidth]{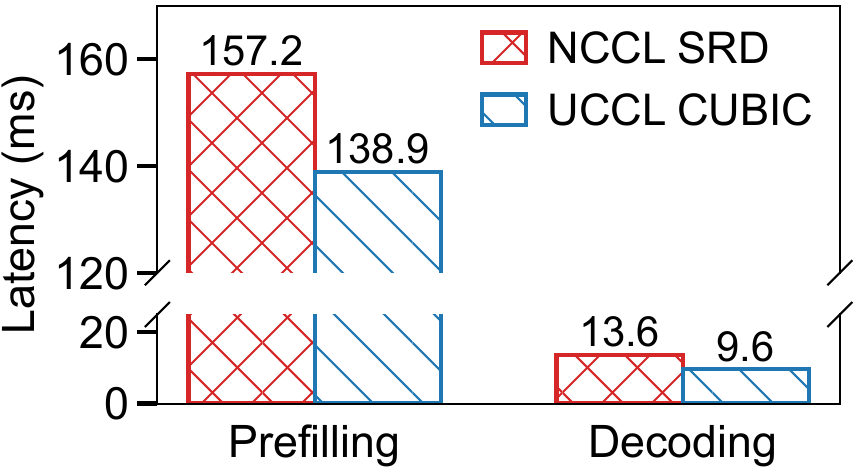}
    \end{minipage}
    \vspace{-0.05in}
    \subcaption{DeepSeek-V3 serving with EP (trace-driven emulation).}
    \label{fig:eval_app_deepseekv3}
\end{minipage}
\postfig
\caption{Application results on EFA (NVLink+SHM disabled).}
\label{fig:application}
\postfigcaption
\end{figure}

Figure~\ref{fig:eval_app_resnet} shows the results for ResNet distributed training. \sysname reduces the training epoch time by 1.07-1.11$\times$ compared to SRD. 
Figure~\ref{fig:eval_app_deepseekv3} shows the results for DeepSeek-V3 serving. \sysname reduces the per-request prefilling and decoding latency by 1.13$\times$ and 1.42$\times$ respectively. 
These experiments demonstrate that \sysname is able to bring significant benefits to real-world ML applications.

\presub
\subsection{\sysname Extensibility}
\postsub\label{ssec:extensibility_eva}

\begin{figure}[!t]
\centering
\begin{minipage}{0.215\textwidth}
\includegraphics[width=\textwidth]
{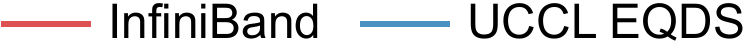}\vspace{0.02in}
\end{minipage}
\par
\prefig
\begin{minipage}{0.23\textwidth}
    \centering
    \includegraphics[width=\linewidth]{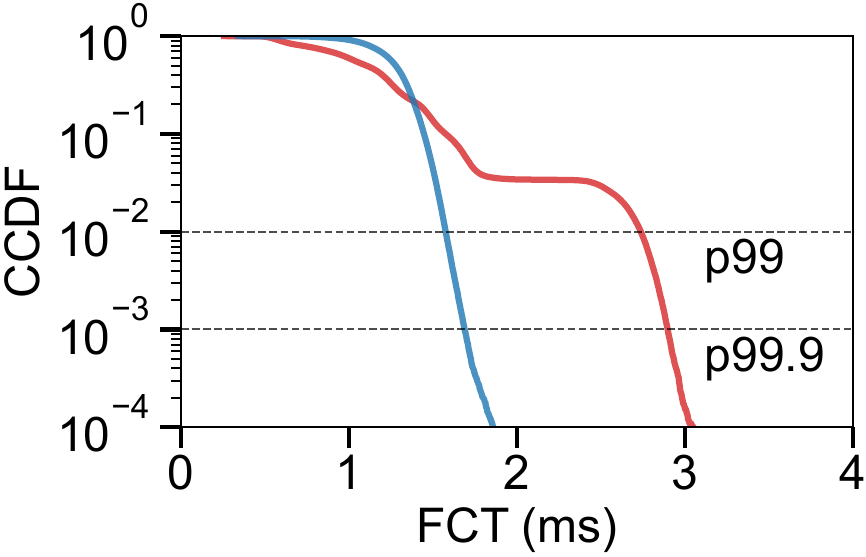}
    \subcaption{Incast traffic.}
    \label{fig:eval_incast_incast_ccdf}
\end{minipage}
\hfill
\begin{minipage}{0.23\textwidth}
    \centering
    \includegraphics[width=\linewidth]{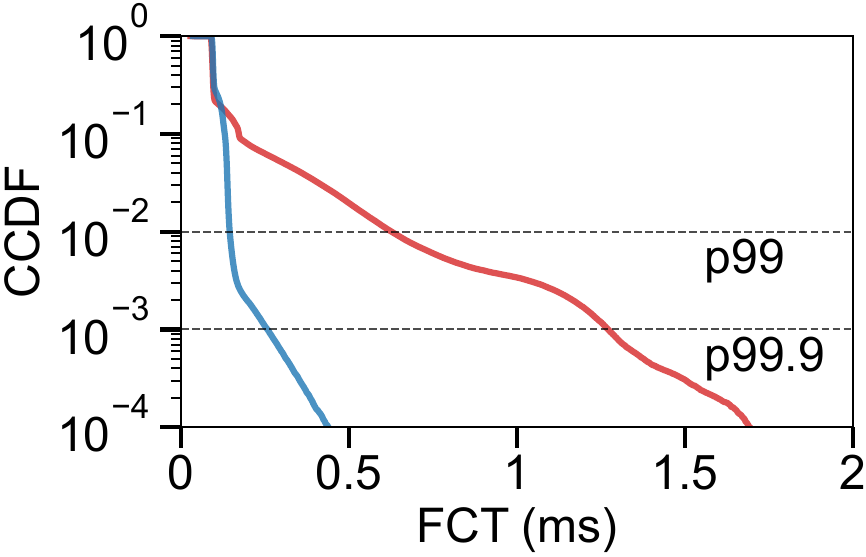}
    \subcaption{Permutation traffic.}
    \label{fig:eval_incast_pt_ccdf}
\end{minipage}
\postfig
\caption{Complementary CDF of FCT (Flow Completion Time) on \cx when co-locating 15-to-1 incast and permutation traffic.}
\label{fig:incast}
\postfigcaption
\end{figure}

\subsubsection{Handling network incast}
In a shared RDMA network, network incast can lead to victim flows due to PFC propagation~\cite{srnic_nsdi23, pfc_deadlocks, irn}. 
Such network incast can be caused by the Expert Parallelism in MoE serving (\S\ref{ssec:motivate_extend}) and gather/reduce collectives in multi-level parallelism of LLM training~\cite{google_pathway, llama3_tech}. 
This experiment tries to create a scenario where network incast co-exists with other collective traffic, and evaluates how \sysname performs compared to the RDMA hardware transport on \cx. 
We co-locate 15-to-1 incast traffic and 16-NIC permutation traffic~\cite{dc_mptcp, lb_routing, strack, smartt}, where each NIC streams data to another and no NIC receives more than one stream, representing typical collectives. 
For both types of traffic, each NIC sends 1MB messages with at most four in flight. 
We compare the receiver-driven EQDS in \sysname vs. the sender-driven InfiniBand CC in ConnectX-7. 

Figure~\ref{fig:eval_incast_incast_ccdf} and~\ref{fig:eval_incast_pt_ccdf} show the FCT distributions for incast traffic and permutation traffic, respectively. 
Compared to InfiniBand, \sysname EQDS reduces P99/P99.9 latency by 1.73$\times$/1.72$\times$ for incast traffic and 4.50$\times$/4.88$\times$ for permutation traffic. 
This is because the InfiniBand CC only reduces the sending rate when severe congestion and queue build-up have occurred on the incast switch port; but this is too late, as the Credit-Based Flow Control~\cite{ib_cc} (PFC equivalent in InfiniBand) has already paused all upstream ports and heavily disturbed victim flows (\ie, permutation traffic). 
Conversely, \sysname EQDS proactively controls the rate of all senders on the receiver side, which reduces queue build-up and avoids upstream ports entering the pausing state. 

\presubsub
\subsubsection{Handling packet loss}
We then evaluate how \sysname selective retransmission handles packet loss, compared to hardware-based loss recovery. 
Furthermore, \sysname coalesces transport decisions like loss recovery at chunk granularity for high software efficiency; therefore, we also test whether this design choice would cause poor loss recovery performance. 
To this end, we instrument different packet loss rates in software when running NCCL collectives over \sysname UC. 
Our packet loss instrumentation has considered the chunking nature in \sysname, where any packet loss will cause the whole chunk loss.
Unfortunately, we cannot instrument packet loss for our \cx NICs as that would require reconfiguring the NIC or switch behaviors, both of which we do not have access to. Instead, we cite comparable numbers from prior literature Flor~\cite{flor}. 

\begin{figure}[!t]
\centering
\includegraphics[width=0.44\textwidth]
{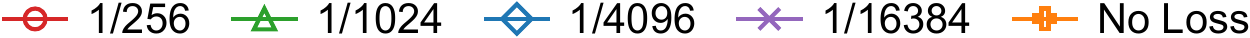}
\vspace{0.05in}
\prefig
\hfill

\begin{minipage}{0.48\textwidth}
\centering
    \begin{minipage}{0.60\textwidth}
        \centering
        \includegraphics[width=\linewidth]{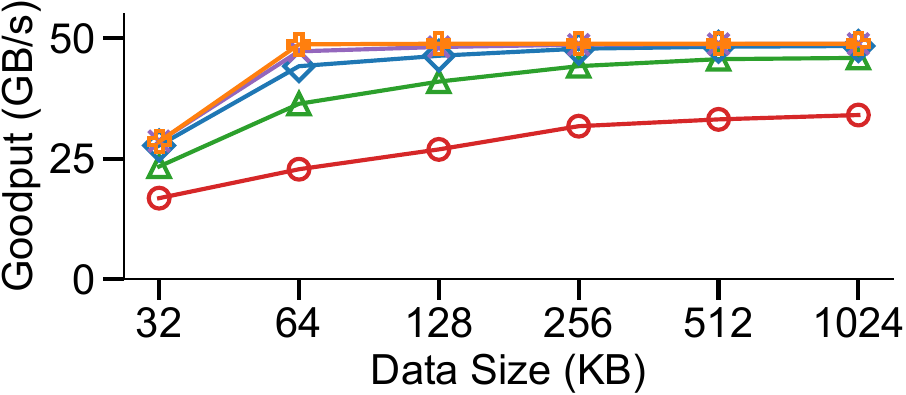}
        \postfig\vspace{-0.1in}
    \end{minipage}
\end{minipage}
\vspace{-0.10in}
\caption{Goodput of \sysname under different packet drop ratios. \sysname implements selective retransmission with a dynamic RTO threshold (around several RTTs).}
\label{fig:eval_pkt_loss}
\postfigcaption\vspace{0.05in}
\end{figure}

We pick two GPUs from the AMD testbed, each using one NIC for communication. 
The client NIC establishes a connection with the server NIC and keeps 16 inflight messages. 
Similar to Flor~\cite{flor}, we use a single QP for the connection, and disable congestion control to avoid sending rate backoff. 
We vary the message size (32KB$\sim$1MB) and measure the goodput under different packet drop ratios. 
Figure~\ref{fig:eval_pkt_loss} shows the results. \sysname has only $\sim$1\% performance drop under 1/16384, 1/4096 drop ratio, compared to the reported 26\%$\sim$42\% drop of RDMA hardware transport in Flor~\cite[Figure 7]{flor}. Even under a high drop ratio of 1/1024 and 1/256, \sysname only experiences a performance drop of 6\%$\sim$30\%, compared to the reported 59\%$\sim$76\% drop in RDMA hardware transport. 

\presub
\subsection{\sysname Scalabality}
\postsub\label{ssec:scalability}

\subsubsection{Varying chunk size}
\label{sssec:chunk_size}
\begin{figure}[!t]
\centering
\prefig
\begin{minipage}{0.23\textwidth}
    \centering
    \includegraphics[width=0.80\textwidth]{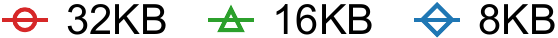} \\[0.05in]
    \includegraphics[width=\textwidth]{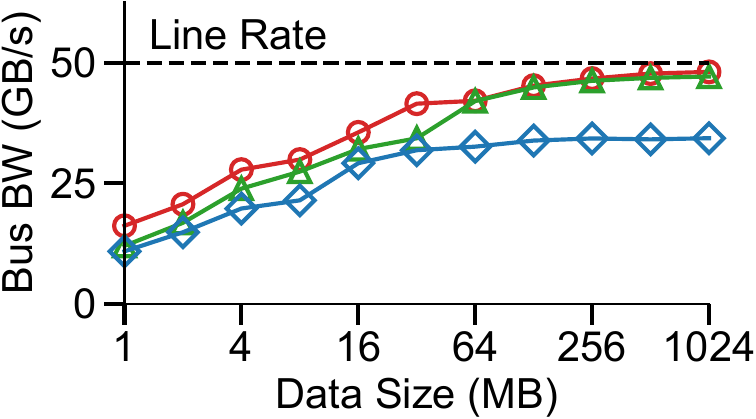}
    \subcaption{Varying chunk size.}
    \label{fig:eval_chunk_size}
\end{minipage}
\hfill
\begin{minipage}{0.23\textwidth}
    \centering
    \includegraphics[width=0.90\textwidth]
    {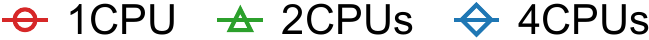} \\[0.05in]
    \includegraphics[width=\textwidth]{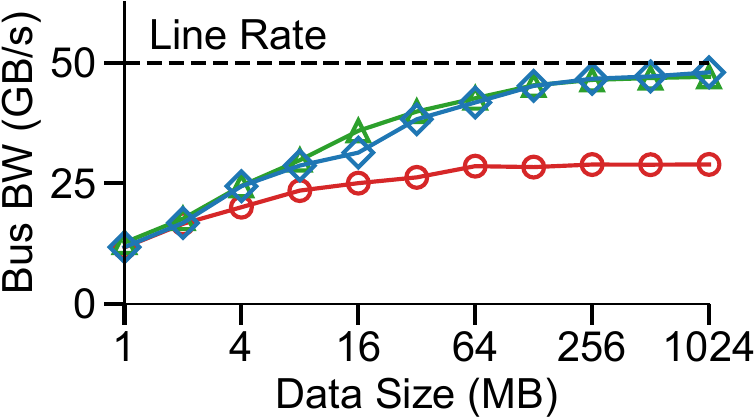}
    \subcaption{Varying CPU cores.}
    \label{fig:eval_asic_cpu}
\end{minipage}
\postfig
\caption{\sysname \alltoall on \cx with different chunk sizes and number of CPU cores per NIC (NVLink+SHM disabled). }
\label{fig:eval_chunk_size_asic_cpu}
\postfigcaption
\end{figure}

Figure~\ref{fig:eval_chunk_size} shows the \alltoall performance on the \cx testbed when varying chunk sizes. 
Our results reveal that saturating the line rate requires a somewhat medium chunk size, \ie, $\ge$16KB or 4 MTU-sized chunks. 
This experiment demonstrates the performance benefits of control coalescing for \sysname software transport. 

\presubsub
\subsubsection{Varying number of CPU cores} \label{sssec:cpu_scalability}
Figure~\ref{fig:eval_asic_cpu} shows the \alltoall performance when varying the number of \sysname engines/CPU cores per NIC. 
For ASIC-based NICs with segmentation and reassembly offloading, even though one CPU core can only saturate bidirectional 29GB/s, two CPU cores are enough to saturate the full 50GB/s line rate. 
The slight performance drop for 16MB when switching from ``2CPUs'' to ``4CPUs'' should be caused by runtime variability. 
This experiment confirms the high efficiency of \sysname software transport, \ie, 1 CPU core saturating 400G unidirectional traffic (2 CPU cores for 400G bidirectional). 

We show more scalability studies in Appendix~\ref{ssec:more_scalability_exp} when varying numbers of CPU cores, GPUs, and paths on EFA. 

\presub
\subsection{Design Drill-Down}
\postsub\label{ssec:design_drill}

\subsubsection{Connection splitting}\label{sssec:conn_splitting}
We now evaluate how connection splitting impacts collective performance in \sysname. 
Without connection splitting, each \sysname connection is mapped to a single engine out of the two \sysname engines per NIC; with splitting, each \sysname connection dispatches and load balances its messages among the two \sysname engines. 
Our experiment results show that \sysname without connection splitting only achieves 45.7 and 39.9 GB/s max busbw for \allreduce and \alltoall on \cx respectively, while \sysname with splitting achieves 48.9 and 48.5 GB/s respectively, saturating the line rate. 
This is because connection splitting enables multiple CPU cores to process inflight messages in a load-balanced manner. 
We anticipate connecting splitting will become more critical as the NIC scales to 800 Gbps~\cite{nvidia_800g}.

\begin{table}[t]
\prefig
\begin{center}
{\footnotesize
\setlength{\tabcolsep}{2.5pt}
\begin{tabular}{ccccc}
    \toprule
    \makecell{Delay ($\mu s$)}
    &\makecell{\uallreduce\\ Light}
    &\makecell{\ualltoall\\ Light}
    &\makecell{\uallreduce\\ Heavy}
    &\makecell{\ualltoall\\ Heavy}
    \\\toprule
    CC decision P50 & 1.7 & 1.7 & 2.1 & 2.9 \\
    CC decision P99 & 3.6 & 3.4 & 7.0 & 10.8 \\
    ACK turnaround P50 & 2.0 & 3.0 & 4.0 & 7.0 \\
    ACK turnaround P99 & 3.0 & 10.0 & 10.0 & 36.0 \\\toprule
\end{tabular}
}
\end{center}
\caption{CC decision delay and ACK turnaround delay on \cx. ``Light'' uses 1KB-64KB messages, while ``Heavy'' uses 1GB messages. }
\postfigcaption
\label{tab:cc_fidelity}
\end{table}

\presubsub
\subsubsection{Congestion signal fidelity in software}\label{sssec:eval_cc_signal}
\sysname makes a fundamental shift of moving transport from hardware to software. 
This experiment evaluates how much software transport sacrifices in terms of congestion signal fidelity. 
We look at the two important metrics as mentioned in \S\ref{ssec:cc_signal}: CC decision delay and ACK turnaround delay. 
Table~\ref{tab:cc_fidelity} shows the results. 
At light loads, the two delays are limited to at most 10$\mu s$ in both P50 and P99; at high loads, they grow up to 36$\mu s$ for P99. 
Both are on par with typical datacenter RTTs of 10-40 $\mu s$~\cite{aquila_nsdi22}. 
Therefore, this experiment confirms that \sysname software transport has high enough CC fidelity to make precise per-RTT transport decisions~\cite{hpcc, falcon_slides}. 

We show more design drill-downs in Appendix~\ref{ssec:more_design_drilldown}, including the impact of chunk size, LB policy, kernel fusion (for scattered memcpy), and PCIe overhead analyses. 

\presec
\section{Discussion}
\postsec
\label{sec:discussion}

\para{HW-SW interface.}
\sysname works around many challenges to support extensible transport with efficient multipathing on existing RDMA NICs, at the cost of QP swapping overhead (though not high for ML collectives), control coalescing, and more. 
\sysname performance and control granularity would get further boosted if there were a better HW-SW interface in RDMA NICs. 
We would like to highlight a few points based on our experience of developing \sysname: 
(1) UC abstraction is powerful with segmentation and reassembly offloading, but using many UC QPs for multipathing incurs some degree of QP swapping overhead; instead, it should evolve into a multipath UC abstraction that allows the software to specify different flow entropies for different verbs over a single QP, just like how \sysname specifies different UDP ports for AF\_XDP. 
(2) NIC HW should expose more congestion signals to SW, such as ECN marks and packet trimming information; these signals can be embedded into CQEs just like hardware timestamps. 

\para{GPU-driven communication} like DeepEP~\cite{deepep} leverages NVIDIA IBGDA~\cite{ibgda} to issue RDMA verbs directly from GPUs to the RDMA NICs. 
Although seemingly conflicted with \sysname's CPU-driven designs, GPU-driven communication could still be made compatible with \sysname. 
The key is to leverage the IBGDA CPU-assisted mode~\cite{ibgda_cpu_assisted}, where the GPU forwards RDMA requests to a CPU proxy that issues RDMA verbs; with this, we could implement \sysname's extensible transport layer inside the CPU proxy. 
The trade-off is on performance: NVIDIA has reported that this CPU-assisted IBGDA would sacrifice 10\% performance compared to the traditional IBGDA~\cite{ibgda_cpu_assisted}. 
We intend to leave the integration and performance enhancement of CPU-assisted IBGDA + \sysname as future work. 

\presec
\section{Other Related Work}
\label{sec:relatedwork}
\postsec

Recent work SCR~\cite{whitebox_rdma} implements receiver-driven CC and multipathing on the DPA (DataPath Accelerator, essentially 16 RISC-V cores) of NVIDIA BlueField-3 SmartNICs. 
It would have similar performance issues as AWS EFA NICs. 
Moreover, DPA programmability is limited to what the NIC hardware supports, \eg, only rate-based control, while packet reliability and retransmission are still baked into the hardware. 
Because of this, SCR needs to alter the original receiver-driven CC to let credits represent available bandwidth rather than bytes. 
For multipathing, SCR only demonstrates two paths; given the limited L1/L2 cache in DPA~\cite{whitebox_rdma}, it is not clear if SCR could scale to hundreds of paths~\cite{strack, smartt, eqds}. 
 
Google GPUDirect-TCPX~\cite{devmem_tcp} integrates GPUDirect into the kernel TCP stack by leveraging the Header-Data Split feature in certain non-RDMA NICs. 
Instead, \sysname targets both RDMA and non-RDMA NICs, and further supports efficient multipathing. 
C4~\cite{c4_alibaba} does coarse-grained flow-level traffic planning and path selection in LLM training, without programmability for low-level RDMA transport components such as CC and loss recovery. 
MSCCL~\cite{msccl} supports customizing collective communication algorithms and could work together with \sysname. 
Lastly, \sysname is inspired by a line of work that targets extensibility for CPU applications, \eg, Google 1RMA~\cite{1rma}, eRPC~\cite{erpc_nsdi19}, RoGUE~\cite{rogue_socc18}, and IO-TCP~\cite{iotcp} for extensible RDMA transport, and SPIN~\cite{spin_offload}, Exokernel~\cite{exokernel}, and VINO~\cite{vino} for extensible OS. 

\presec
\section{Conclusion}
\postsec
\label{sec:conclusion}

\sysname is an extensible and efficient software transport layer for GPU networking. 
It achieves network extensibility by separating the control path and data path for existing RDMA NICs and running the transport control path in software. 
Meanwhile, it achieves hardware-level performance by leveraging techniques like control coalescing and connection splitting. 
We hope \sysname opens the door to the productization of new research proposals on network transports for ML workloads. 
\sysname is open-sourced at \href{https://github.com/uccl-project/uccl}{https://github.com/uccl-project/uccl}. 

{
\bibliographystyle{./templates/ACM-Reference-Format}
\bibliography{ref}
}

\newpage
\appendix

\presub
\section{Interface to Extensibility}
\postsub
\label{ssec:interface}

By executing control decisions in software, \sysname allows flexibly extending its transport implementation for different scenarios. 
To ease the development of new multipath transport such as new CC or LB policies among different paths, \sysname exposes a set of expressive interfaces to collective library or ML application developers, as shown in Listing~\ref{list:apis}. 
\begin{itemize}
\item \texttt{onChunkSize} is called when \sysname chunks a message for transmission, and it returns the permitted chunk size for now. CC could enforce window control here. After it returns, \sysname will build a \texttt{chunk\_desc} for the chunk. 
%
\item \texttt{onPacingChunk} determines if a chunk needs to queue in the timing wheel for rate pacing, and returns true if so.
\item \texttt{onSelectPath} is called when a chunk is ready for transmission. \texttt{conn\_state} contains rich information for selecting path, \eg, RTT scoreboard for each path. It returns the selected \texttt{path\_id} (\ie, QP ID) for transmission. 
\item \texttt{onTxRtxChunk} is called when the reliable transport wants to retransmit a chunk, and it returns true if this chunk is permitted to retransmit. CC could enforce window control for the retransmitted chunk here. 
\item \texttt{onRxChunk} is called when receiving a data chunk. 
\item \texttt{onRxRtxChunk} is called when receiving a retransmitted chunk. CC could react to the retransmission chunk here. 
\item \texttt{onRxACK} is called when receiving an ACK. CC could react to the ACK here.
\item \texttt{onRxCredit} is called when receiving a credit. Receiver-driven CC could react to the credit here.
\end{itemize}

\begin{lstlisting}[language=C++, style=myStyle, captionpos=b, float=htbp,
    caption={\sysname interface to extending multipath transport.},
    label={list:apis}, abovecaptionskip=0.10in, belowcaptionskip=-0.10in]
func onChunkSize(conn_state, remaining_bytes) -> chunk_sz;
func onPacingChunk(conn_state, chunk_desc) -> pacing_or_not;
func onSelectPath(conn_state, chunk_desc) -> path_id;
func onTxRtxChunk(conn_state, chunk_desc) -> rtx_or_not;
func onRxChunk(conn_state, ctrl_hdr);
func onRxRtxChunk(conn_state, ctrl_hdr);
func onRxACK(conn_state, sack_hdr);
func onRxCredit(conn_state, credit_hdr):
\end{lstlisting}

\section{EQDS Implementation Under \sysname}
\label{sec:eqds_impl}

\begin{figure}[!t]
\centering
\prefig
\begin{minipage}[t!]{0.475\textwidth}
\includegraphics[width=\textwidth]{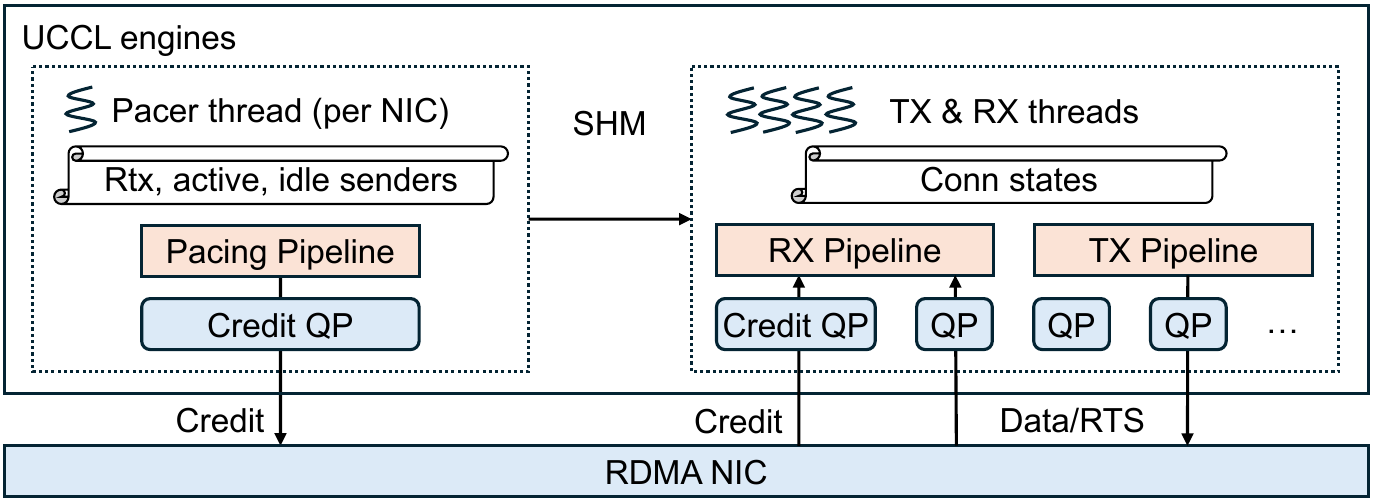}
\end{minipage}
\postfig
\caption{Implementing receiver-driven CC under \sysname. }
\label{fig:eqds}
\postfigcaption\vspace{0.1in}
\end{figure}

Figure~\ref{fig:eqds} shows the overall implementation. 
For each NIC, \sysname creates a dedicated pacer thread that operates at a constant rate (derived from the NIC bandwidth) to select candidate senders, allocate credits, and send credits following the EQDS algorithm. 
Each pacer thread uses a credit UD QP for sending credit packets, and each TX\&RX thread also has its own credit QP for receiving credit packets from remote pacers. 
The pacer thread maintains three lists, \ie, rtx (retransmission), active, and idle sender list, with priorities from high to low. 
After the TX\&RX threads receive data chunks, they will notify the pacer thread via an efficient atomic write in the SHM. 
Then the pacer will update the sender list: 
senders who encounter packet loss will be put into the rtx list; 
senders who have satisfied their requirements will be put into the idle list; 
otherwise, they will be put into the active list. 
%
%
One note is that as packet tramming is not available in our RDMA NICs and switches, \sysname uses timeout + RTS (Request-To-Send) as a replacement, as suggested by the EQDS paper. 

\section{More Evaluation Results}
\label{sec:more_eval}

\subsection{More collectives in ML workloads}
\label{ssec:more_coll}

\subsubsection{Allgather and \reducescatter}
\begin{figure}[!t]
\centering
\includegraphics[width=0.24\textwidth]
{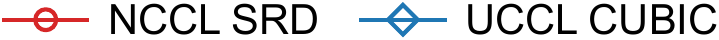}\vspace{0.05in}
\prefig
\begin{minipage}{0.48\textwidth}
    \begin{minipage}{0.48\textwidth}
        \centering
        \includegraphics[width=\linewidth]{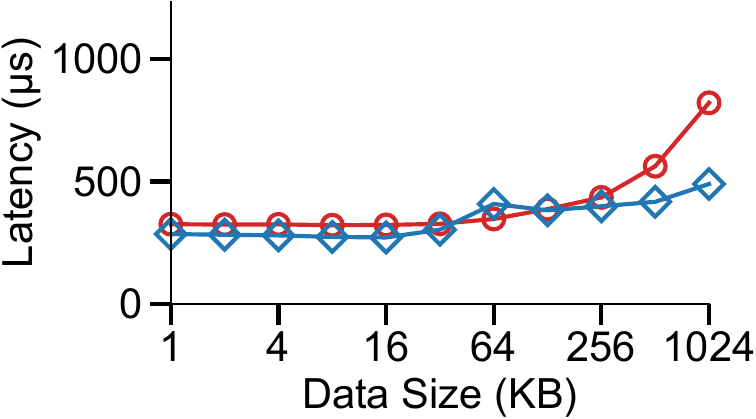}
        \label{fig:EFA_allgather_Latency}
    \end{minipage}
    \hfill
    \begin{minipage}{0.48\textwidth}
        \centering
        \includegraphics[width=\linewidth]{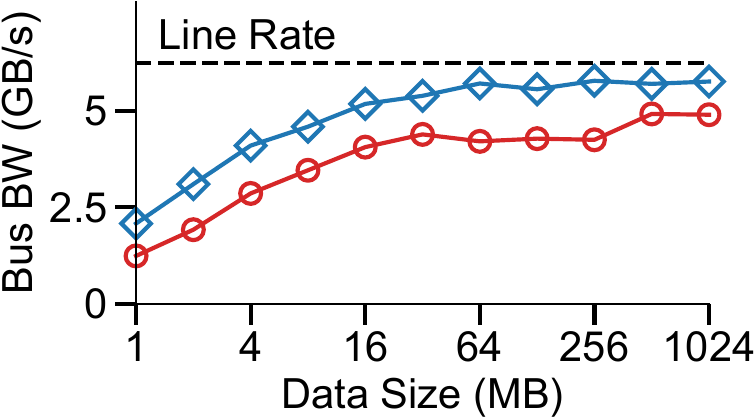}
        \label{fig:EFA_allgather_BusBandwidth}
    \end{minipage}
    \postfigdouble
    \subcaption{\uallgather.}\label{fig:more_coll_ag}
\end{minipage}
\hfill
\begin{minipage}{0.48\textwidth}
    \begin{minipage}{0.48\textwidth}
        \centering
        \includegraphics[width=\linewidth]{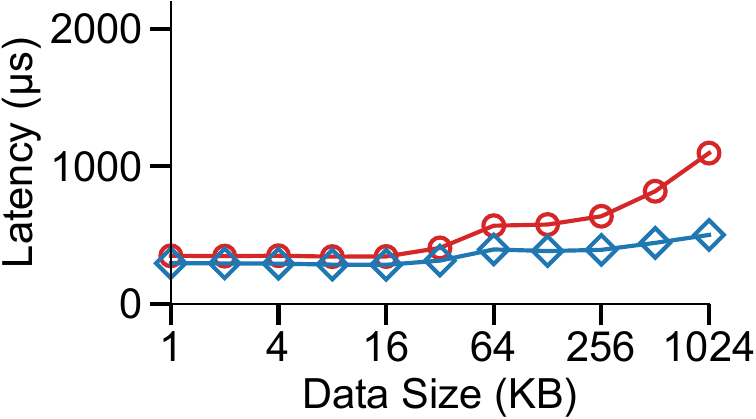}
        \label{fig:EFA_reducescatter_Latency}
    \end{minipage}
    \hfill
    \begin{minipage}{0.48\textwidth}
        \centering
        \includegraphics[width=\linewidth]{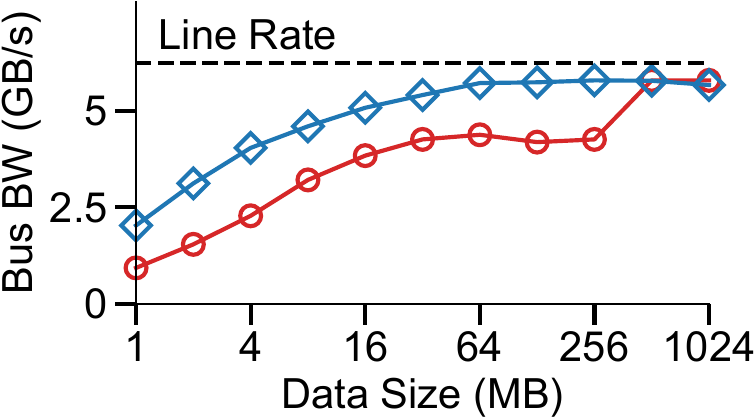}
        \label{fig:EFA_reducescatter_BusBandwidth}
    \end{minipage}
    \postfigdouble
    \subcaption{\ureducescatter.}\label{fig:more_coll_rs}
\end{minipage}
\vspace{-0.15in}
\caption{NCCL-tests results on EFA (NVLink+SHM disabled to simulate a larger testbed).}
\label{fig:eval_efa_more_coll_2}
\postfigcaption
\end{figure}

These two are representative collectives used in PyTorch FSDP (Fully Sharded Data Parallelism)~\cite{fsdp}, exhibiting low network congestion~\cite{llm_meta}. Figure~\ref{fig:more_coll_ag} and Figure~\ref{fig:more_coll_rs} compare the \allgather and \reducescatter performance of \sysname and SRD on the EFA testbed. \sysname outperforms SRD by up to 1.68$\times$ for \allgather and up to 2.18$\times$ for \reducescatter. 

\subsubsection{Multi-collectives}
\begin{figure}[!t]
\centering
\includegraphics[width=0.24\textwidth]
{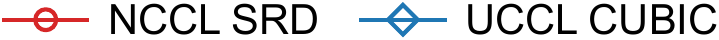}\vspace{0.05in}
\prefig
\begin{minipage}{0.48\textwidth}
    \begin{minipage}{0.48\textwidth}
        \centering
        \includegraphics[width=\linewidth]{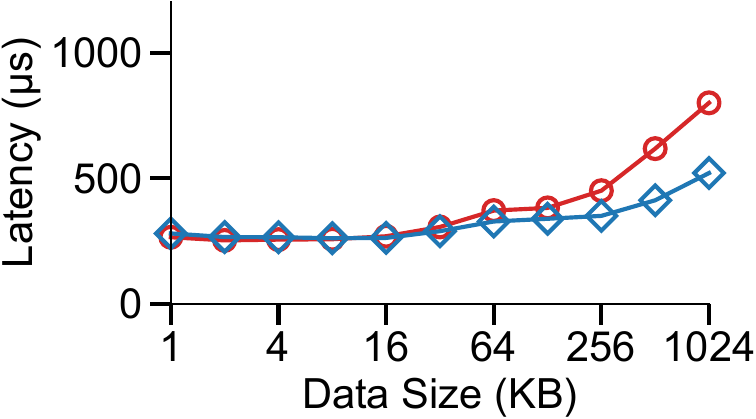}
        \label{fig:EFA_multi_allreduce_Latency}
    \end{minipage}
    \hfill
    \begin{minipage}{0.48\textwidth}
        \centering
        \includegraphics[width=\linewidth]{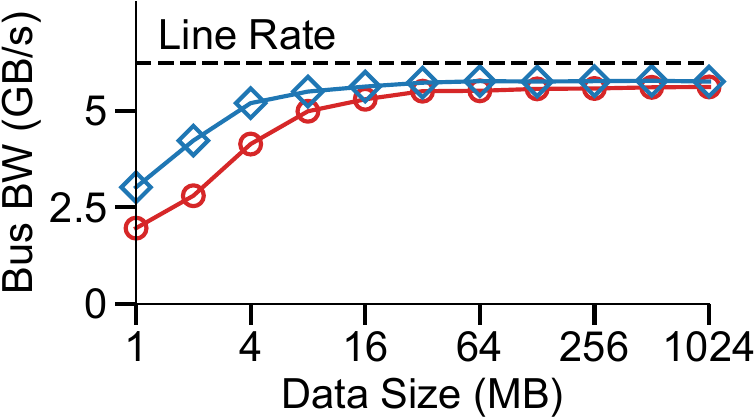}
        \label{fig:EFA_multi_allreduce_BusBandwidth}
    \end{minipage}
    \postfigdouble
    \subcaption{\umallreduce. }\label{fig:more_coll_mar}
\end{minipage}
\hfill
\begin{minipage}{0.48\textwidth}
    \begin{minipage}{0.48\textwidth}
        \centering
        \includegraphics[width=\linewidth]{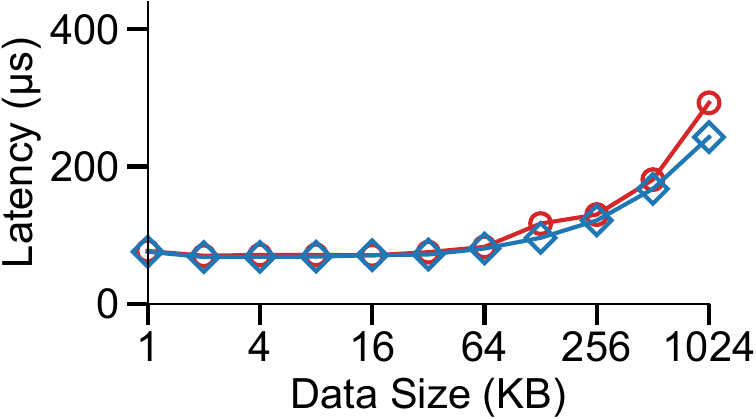}
        \label{fig:EFA_multi_alltoall_Latency}
    \end{minipage}
    \hfill
    \begin{minipage}{0.48\textwidth}
        \centering
        \includegraphics[width=\linewidth]{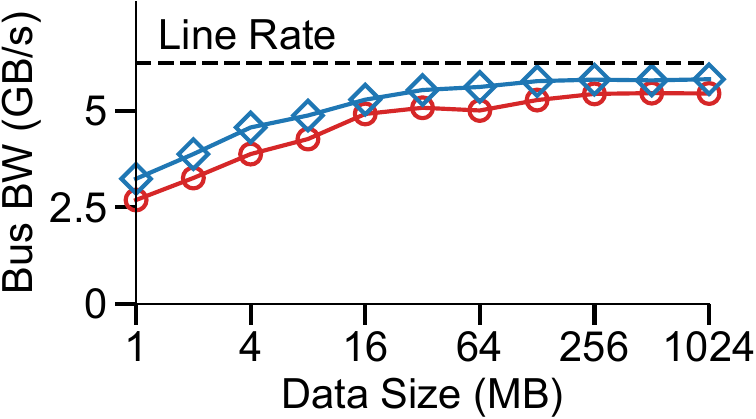}
        \label{fig:EFA_multi_alltoall_BusBandwidth}
    \end{minipage}
    \postfigdouble
    \subcaption{\umalltoall.}\label{fig:more_coll_maa}
\end{minipage}
\hfill
\begin{minipage}{0.48\textwidth}
    \begin{minipage}{0.48\textwidth}
        \centering
        \includegraphics[width=\linewidth]{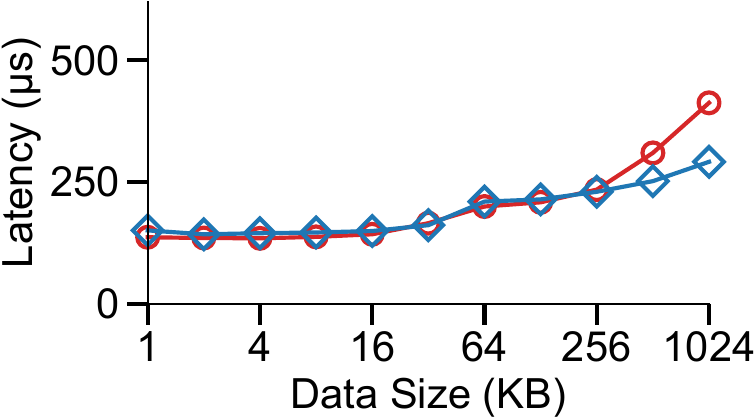}
        \label{fig:EFA_multi_allgather_Latency}
    \end{minipage}
    \hfill
    \begin{minipage}{0.48\textwidth}
        \centering
        \includegraphics[width=\linewidth]{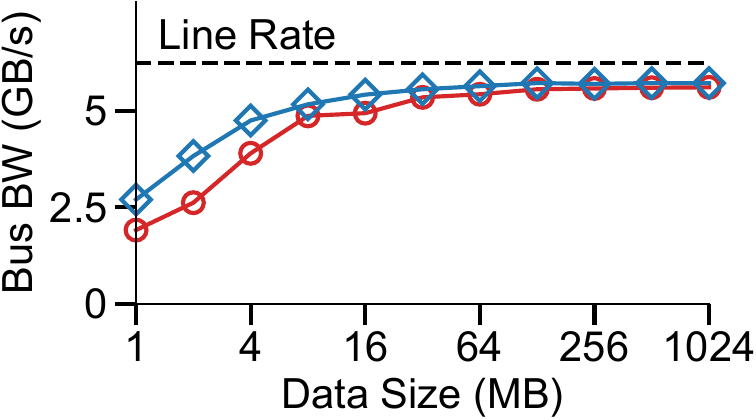}
        \label{fig:EFA_multi_allgather_BusBandwidth}
    \end{minipage}
    \postfigdouble
    \subcaption{\umallgather.}\label{fig:more_coll_mag}
\end{minipage}
\hfill
\begin{minipage}{0.48\textwidth}
    \begin{minipage}{0.48\textwidth}
        \centering
        \includegraphics[width=\linewidth]{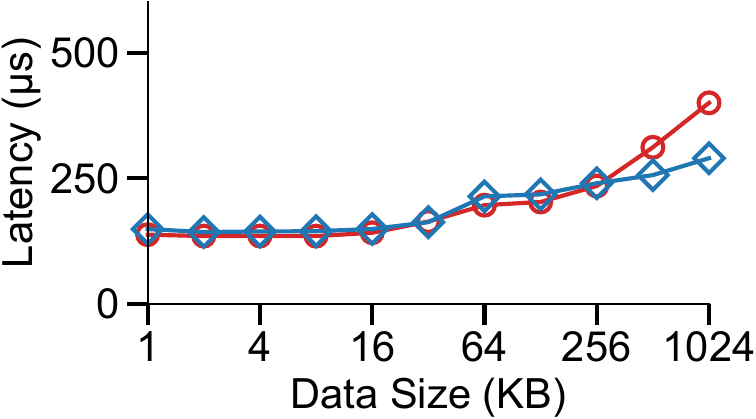}
        \label{fig:EFA_multi_reducescatter_Latency}
    \end{minipage}
    \hfill
    \begin{minipage}{0.48\textwidth}
        \centering
        \includegraphics[width=\linewidth]{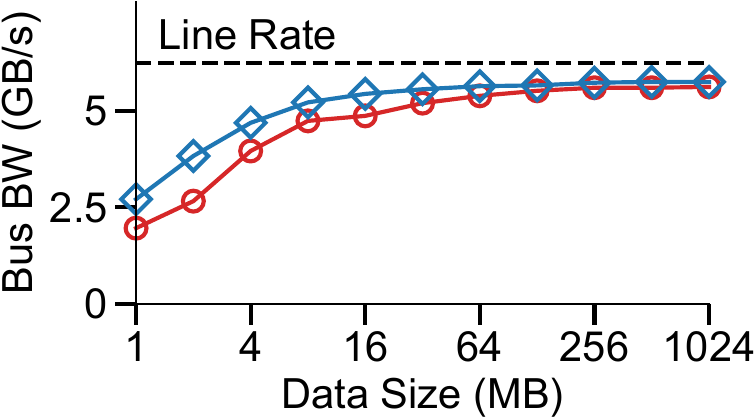}
        \label{fig:EFA_multi_reducescatter_BusBandwidth}
    \end{minipage}
    \postfigdouble
    \subcaption{\umreducescatter.}\label{fig:more_coll_mrs}
\end{minipage}
\vspace{-0.15in}
\caption{Multi-collective results on EFA (NVLink+SHM disabled).}
\label{fig:eval_efa_more_coll_1}
\postfigcaption
\end{figure}

In multi-collectives, GPUs with the same local rank in each server form a collective group, and multiple groups conduct collectives in parallel. For example, \mallreduce is used in ML workloads with intra-server Tensor Parallelism (TP) + inter-server Data Parallelism (DP). 
We evaluate multi-collective performance by setting the environment variable \texttt{NCCL\_TESTS\_SPLIT\_MASK=0x7} in NCCL-tests. 
Figure~\ref{fig:more_coll_mar}, \ref{fig:more_coll_maa}, \ref{fig:more_coll_mag} and \ref{fig:more_coll_mrs} compare the \mallreduce, \malltoall, \mallgather and \mreducescatter performance of \sysname and SRD on the EFA testbed. \sysname outperforms SRD by up to 1.54$\times$, 1.22$\times$, 1.46$\times$, and 1.44$\times$ for \mallreduce, \malltoall, \mallgather and \mreducescatter, respectively. 

\subsection{\sysname AF\_XDP performance}
\label{ssec:afxdp_perf}

\begin{figure}[!t]
\centering
\includegraphics[width=0.245\textwidth]
{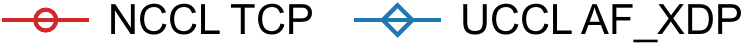}
\prefig
\begin{minipage}{0.48\textwidth}
    \begin{minipage}{0.48\textwidth}
        \centering
        \includegraphics[width=\linewidth]{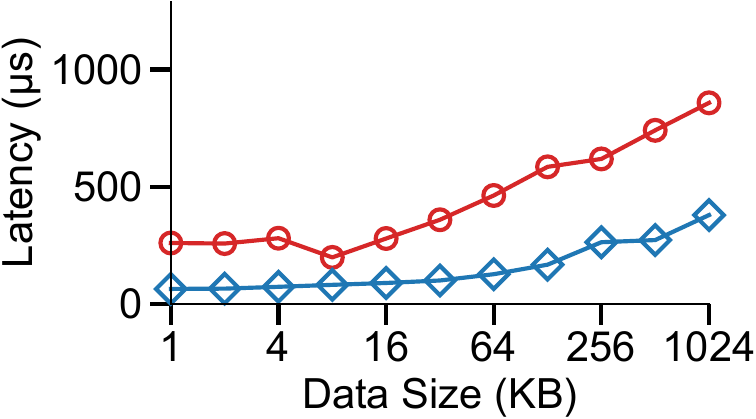}
        \label{fig:latency_allreduce_ena_nic_ppg}
    \end{minipage}
    \hfill
    \begin{minipage}{0.48\textwidth}
        \centering
        \includegraphics[width=\linewidth]{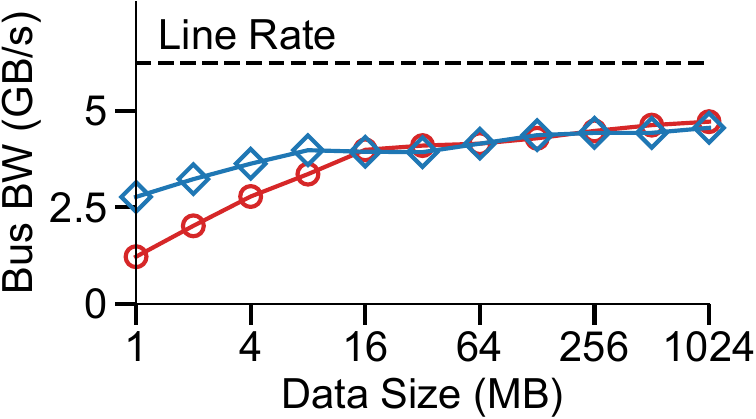}
        \label{fig:bandwidth_allreduce_ena_nic_ppg}
    \end{minipage}
    \vspace{-0.20in}
    \subcaption{With proximity placement group.}
    \label{fig:eval_ena_allreduce_ppg}
\end{minipage}
\hfill
\begin{minipage}{0.48\textwidth}
    \begin{minipage}{0.48\textwidth}
        \centering
        \includegraphics[width=\linewidth]{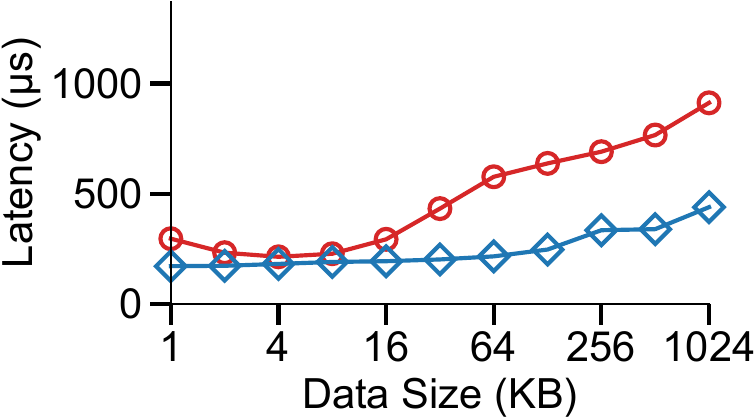}
        \label{fig:latency_allreduce_ena_nic}
    \end{minipage}
    \hfill
    \begin{minipage}{0.48\textwidth}
        \centering
        \includegraphics[width=\linewidth]{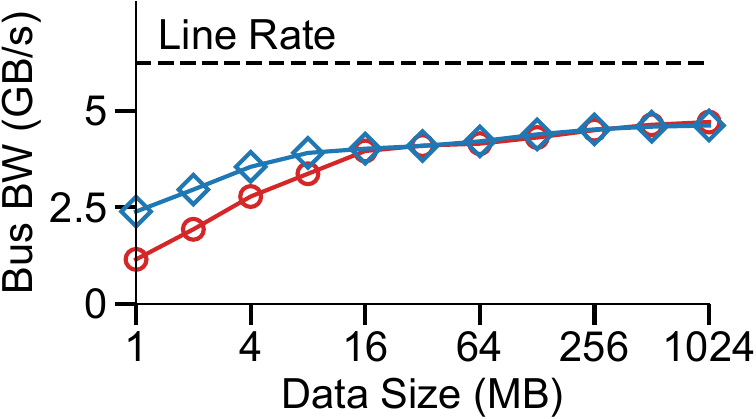}
        \label{fig:bandwidth_allreduce_ena_nic}
    \end{minipage}
    \vspace{-0.20in}
    \subcaption{Without proximity placement group. }
    \label{fig:eval_ena_allreduce}
\end{minipage}
\vspace{-0.15in}
\caption{\uallreduce performance on non-RDMA NICs using two AWS \texttt{g4dn.8xlarge} VMs each with a 50 Gbps AWS ENA NIC.
}
\label{fig:eval_ena}
\postfigcaption\vspace{0.10in}
\end{figure}

Figure~\ref{fig:eval_ena} compares the collective performance of \sysname AF\_XDP vs. NCCL kernel TCP with and without proximity placement group (PPG). 
This experiment is done on AWS with two 50 Gbps VMs connected through AWS ENA NICs that do not support RDMA. We configure NCCL to use multiple TCP connections that achieve the best performance. 
With PPG that gives low network RTTs between VMs, \sysname achieves up to 4.1$\times$/2.3$\times$ higher performance than NCCL for small/large message ranges; without PPG, \sysname achieves up to 2.7$\times$/2.1$\times$ higher performance. 
This huge gain is because \sysname implements an efficient multipath transport atop the fast user-space packet IO technique AF\_XDP, thus saving significant user-kernel context switching overhead and heavy-weight networking stack traversing in kernel TCP. For data sizes exceeding 16MB, both AF\_XDP and TCP experience bottlenecks due to no GPUDirect support.

\subsection{\sysname Scalability}
\label{ssec:more_scalability_exp}

\subsubsection{Varying number of CPU cores on EFA}
\label{sssec:vary_cpu_efa}

\begin{figure}[!t]
\centering
\includegraphics[width=0.275\textwidth]
{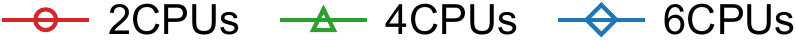}\vspace{0.05in}
\prefig
\begin{minipage}{0.23\textwidth}
    \centering
    \includegraphics[width=\linewidth]{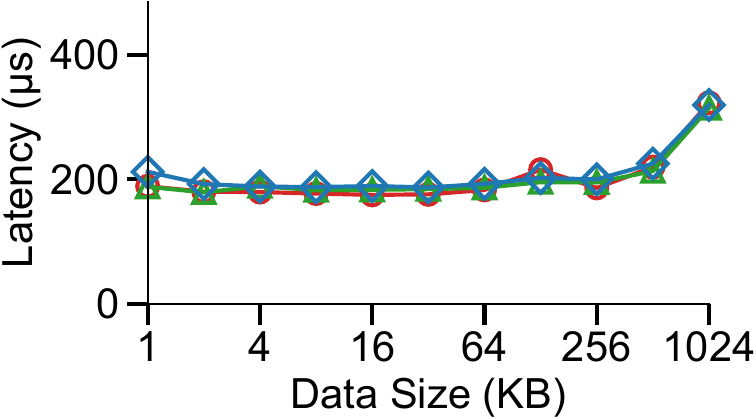}
    \label{fig:latency_alltoall_efa_nic_engine}
\end{minipage}
\hfill
\begin{minipage}{0.23\textwidth}
    \centering
    \includegraphics[width=\linewidth]{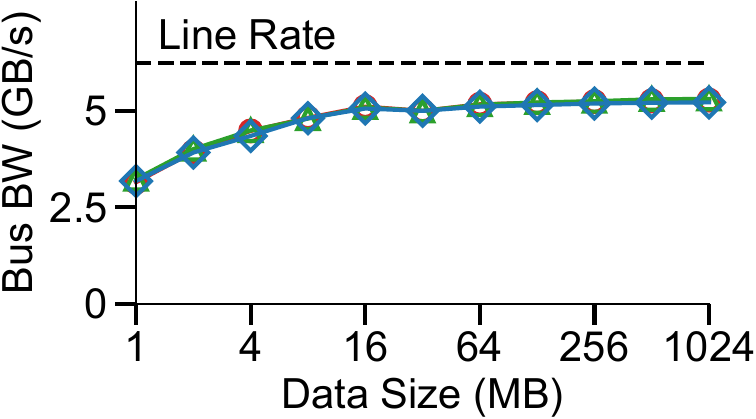}
    \label{fig:bandwidth_alltoall_efa_nic_engine}
\end{minipage}
\postfigdouble
\caption{\sysname \alltoall on EFA with different numbers of CPU cores per NIC (NVLink+SHM disabled). }
\label{fig:eval_efa_cpu}
\postfigcaption\vspace{0.05in}
\end{figure}

Figure~\ref{fig:eval_efa_cpu} shows how the number of CPU cores per NIC impacts \sysname performance on EFA. As mentioned in \S\ref{sssec:cpu_scalability}, 2 cores are sufficient to saturate the EFA line rate, even for the connection-intensive \alltoall collective. 
Overall, thanks to chained posting, \sysname over UD is able to use 1 core to handle 100 Gbps unidirectional traffic on EFA NICs. 

\subsubsection{Varying number of GPUs on EFA}
\label{sssec:vary_gpu_efa}

\begin{figure}[!t]
\centering
\includegraphics[width=0.285\textwidth]
{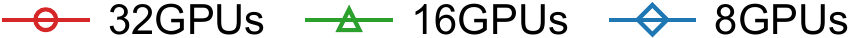}\vspace{0.05in}
\prefig
\begin{minipage}{0.48\textwidth}
    \begin{minipage}{0.48\textwidth}
        \centering
        \includegraphics[width=\linewidth]{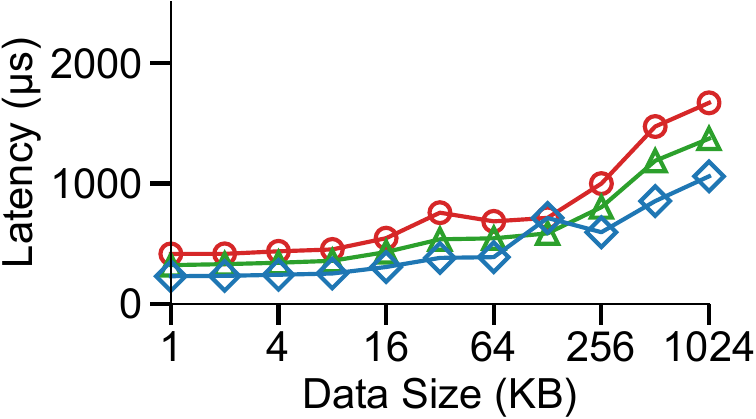}
        \label{fig:latency_allreduce_gpus_efa_nic}
    \end{minipage}
    \hfill
    \begin{minipage}{0.48\textwidth}
        \centering
        \includegraphics[width=\linewidth]{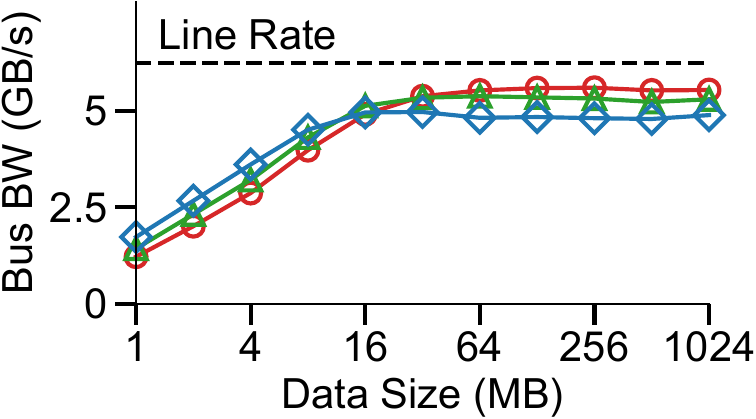}
        \label{fig:bandwidth_allreduce_gpus_efa_nic}
    \end{minipage}
    \postfigdouble
    \subcaption{\uallreduce.}
    \label{fig:eval_efa_gpu_allreduce}
\end{minipage}
\hfill
\begin{minipage}{0.48\textwidth}
    \begin{minipage}{0.48\textwidth}
        \centering
        \includegraphics[width=\linewidth]{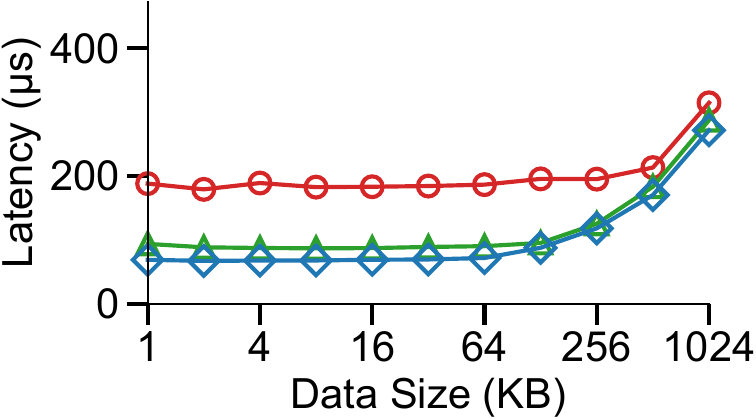}
        \label{fig:latency_alltoall_gpus_efa_nic}
    \end{minipage}
    \hfill
    \begin{minipage}{0.48\textwidth}
        \centering
        \includegraphics[width=\linewidth]{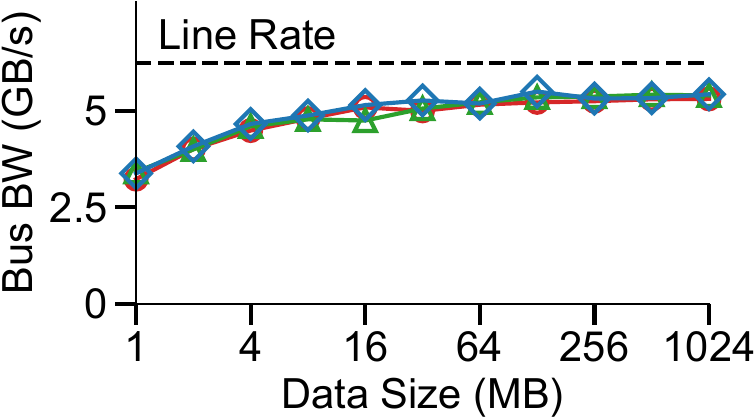}
        \label{fig:bandwidth_alltoall_gpus_efa_nic}
    \end{minipage}
    \postfigdouble
    \subcaption{\ualltoall.}
    \label{fig:eval_efa_gpu_alltoall}
\end{minipage}
\vspace{-0.125in}
\caption{\sysname collectives on EFA with different numbers of GPUs (NVLink+SHM disabled). }
\label{fig:eval_efa_gpu}
\postfigcaption\vspace{0.1in}
\end{figure}

Figure~\ref{fig:eval_efa_gpu} shows how \sysname scales with the number of GPUs on EFA. 
As expected, with less GPUs, \sysname achieves lower latency and higher bus bandwidth. 
But eventually, \sysname is able to approach line rate with $\ge$64MB data size. 
\sysname leverages the connection-less UD on EFA NICs, thus not suffering from QP scalability issues. 

\subsubsection{Varying number of paths on EFA}
\begin{figure}[!t]
\centering
\includegraphics[width=0.43\textwidth]
{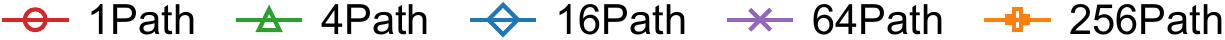}\vspace{0.05in}
\prefig
\begin{minipage}{0.23\textwidth}
    \centering
    \includegraphics[width=\linewidth]{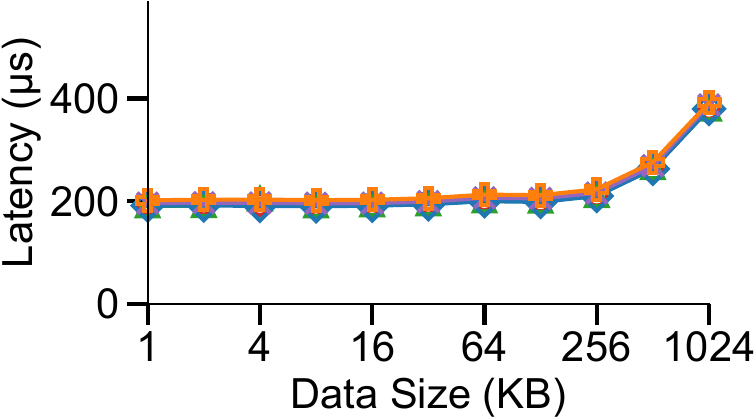}
    \label{fig:latency_alltoall_efa_nic_path}
\end{minipage}
\hfill
\begin{minipage}{0.23\textwidth}
    \centering
    \includegraphics[width=\linewidth]{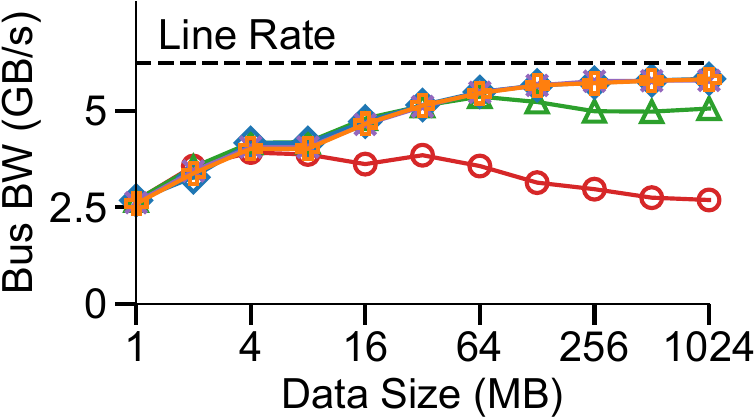}
    \label{fig:bandwidth_alltoall_efa_nic_path}
\end{minipage}
\postfigdouble
\caption{\sysname \alltoall on EFA with different numbers of paths (NVLink + SHM disabled).}
\label{fig:eval_vary_path}
\postfigcaption
\end{figure}

Figure~\ref{fig:eval_vary_path} varies the number of paths on the EFA testbed (which crosses racks with multiple network paths). 
The high-level take here is that multipathing helps mitigate network congestion caused by, \eg, flow collisions. We expect multipath transport to shine more on a larger testbed with more network paths. 

\subsection{Design Drill-Down}
\label{ssec:more_design_drilldown}

\subsubsection{Impact of chunk size and LB policy} \label{sssec:chunk_size}
This experiment aims to study the impact of chunk size and LB policy on the performance of a broad range of ML transports (not just the ones implemented in \sysname). 
To this end, we use a packet-level network simulator \texttt{htsim}~\cite{uec} and implement UEC standard multipath transports based on their papers~\cite{strack, smartt, eqds}. 
We further modify the transport implementation in \texttt{htsim} to vary chunk size and LB policy, \eg, based on ECN or RTT, connection splitting or not. 
We vary the chunk size by changing the MTU size in the simulator without modifying any transport simulation code. 
We vary the LB policy by maintaining per-path ECN/RTT and modifying the code where the transport sender selects the path for each packet. 
Similar to prior work~\cite{dc_mptcp, lb_routing, strack, smartt}, we focus on the permutation traffic pattern to stress-test the transport, where each NIC streams traffic to another and no one receives more than one stream.
We simulate 1024 400G NICs under a fully-provisioned three-tier fattree, and each NIC uses 256 paths to stream 64MB traffic with an ideal completion time of 1.28ms. 

Table~\ref{tab:chunk_lb} shows the permutation traffic completion time under different designs. 
Overall, using 32KB chunk size and switching to RTT degrade transport performance by 17.9\% for sender-driven transport and only 2.8\% for receiver-driven, while connection splitting does not degrade. 
With a 16KB chunk size, the performance degradation of sender-driven transport becomes only 4.1\%. 
Receiver-driven performs better because EQDS leverages in-switch packet trimming to quickly detect network congestion and react~\cite{ndp, eqds}. 
Overall, this experiment shows that control coalescing indeed causes transport performance degradation, but the degradation is moderate in most cases; meanwhile, connection splitting has a negligible impact on transport performance. 

\begin{table}[!t]
\prefig
\begin{center}
{\footnotesize
\begin{tabular}{lll}
    \toprule
    \makecell[l]{Permulation traffic\\ completion time (ms)}
    &\makecell[l]{Sender-driven}
    &\makecell[l]{Receiver-driven}\\\toprule
    4KB + ECN (UEC default)  & 1.45 & 1.43 \\
    32KB + ECN & 1.62 (+11.7\%) & 1.47 (+2.8\%) \\
    32KB + RTT & 1.71 (+17.9\%) & 1.47 (+2.8\%) \\
    32KB + RTT + ConnSplit & 1.70 (+17.2\%) & 1.47 (+2.8\%) \\
    16KB + RTT + ConnSplit & 1.51 (+4.1\%) & 1.42 (-0.7\%) \\
    \toprule
\end{tabular}
}
\end{center}
\caption{Impact of chunk size and LB policy. ``ConnSplit'' means connection splitting, which first selects the least loaded engine and then selects the least loaded path from that engine's paths/QPs using Power-of-Two sampling (\S\ref{ssec:case_multipath}). }
\postfigcaption\vspace{-0.1in}
\label{tab:chunk_lb}
\end{table}


\subsubsection{Impact of kernel fusion.}
%
This experiment aims to quantify the overhead of scattered memcpy and justify \sysname's design choice of employing kernel fusion over kernel launching (\S\ref{ssec:sw_hw_sep}). 
%
In the kernel launching approach, \sysname launches a dedicated copy thread (in CPU) for each \sysname engine at boot time; after receiving all packets of a transport buffer, the engine notifies its associated copy thread via a shared-memory queue; then the copy thread launches a GPU kernel to perform scattered memcpy asynchronously (this kernel is different from the reduction kernel). 
We further optimize the kernel launching overhead with adaptive batching~\cite{ix_osdi14}. 
Note that we cannot run the GPU kernel persistently, as this would cause deadlock with the commonly-used \texttt{cudaDeviceSynchronize()}. 
We also emulate the performance with no scattered memcpy by skipping packet reassembly entirely and disabling data correctness checks in NCCL-tests. 
 
\begin{figure}[!t]
\centering
\includegraphics[width=0.41\textwidth]
{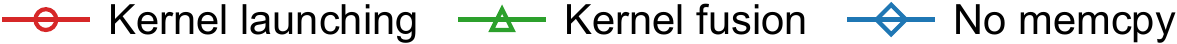}\vspace{0.05in}
\prefig
\begin{minipage}{0.48\textwidth}
    \begin{minipage}{0.48\textwidth}
        \centering
        \includegraphics[width=\linewidth]{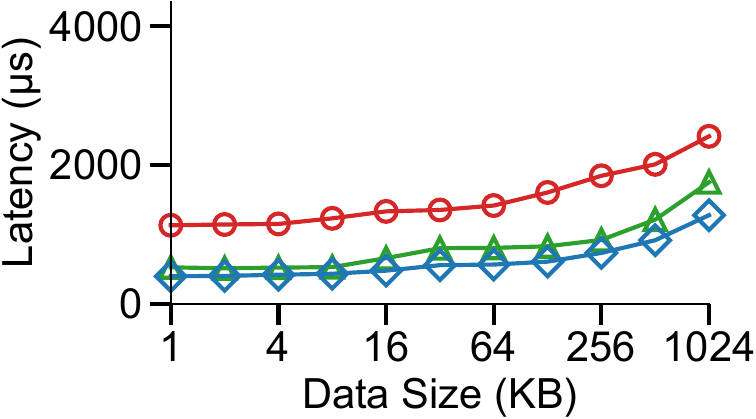}
        \label{fig:EFA_fuse_allreduce_Latency}
    \end{minipage}
    \hfill
    \begin{minipage}{0.48\textwidth}
        \centering
        \includegraphics[width=\linewidth]{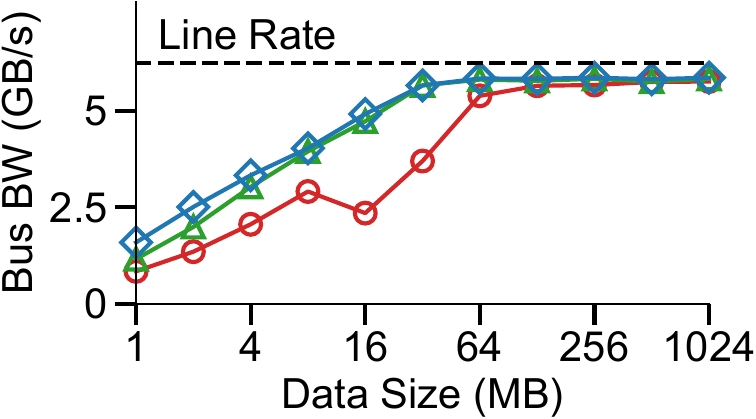}
        \label{fig:EFA_fuse_allreduce_BusBandwidth}
    \end{minipage}
    \postfigdouble
    \subcaption{\uallreduce. 
    }
    \label{fig:eval_efa_fuse_allreduce}
\end{minipage}
\hfill
\begin{minipage}{0.48\textwidth}
    \begin{minipage}{0.48\textwidth}
        \centering
        \includegraphics[width=\linewidth]{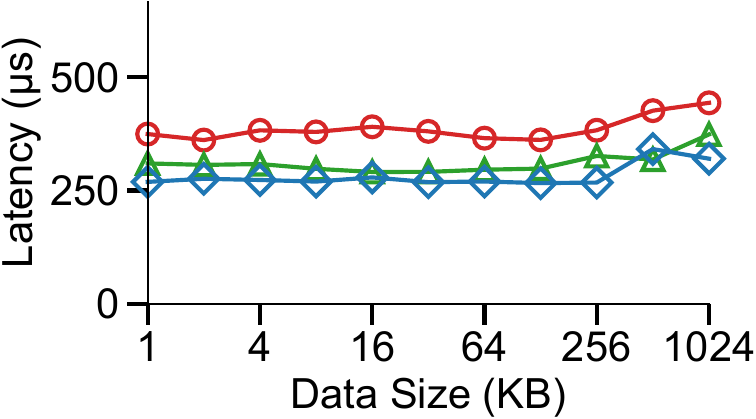}
        \label{fig:EFA_fuse_alltoall_Latency}
    \end{minipage}
    \hfill
    \begin{minipage}{0.48\textwidth}
        \centering
        \includegraphics[width=\linewidth]{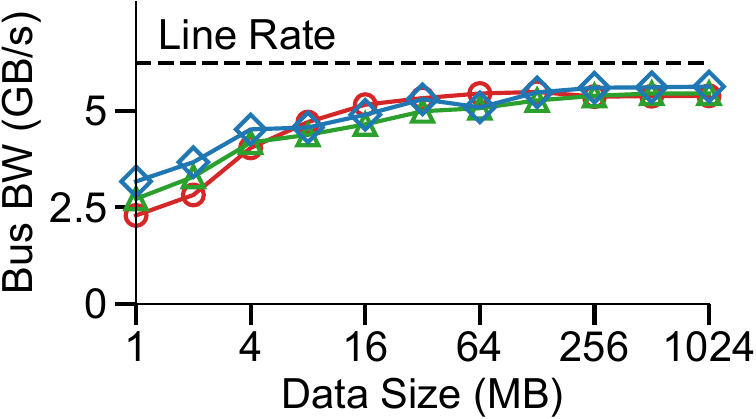}
        \label{fig:EFA_fuse_alltoall_BusBandwidth}
    \end{minipage}
    \postfigdouble
    \subcaption{\ualltoall.}
    \label{fig:eval_efa_fuse_alltoall}
\end{minipage}
\vspace{-0.15in}
\caption{NCCL-tests results on EFA (NVLink+SHM disabled).}
\label{fig:eval_kernel}
\postfigcaption
\end{figure}

Figure~\ref{fig:eval_kernel} shows the collective performance. 
Scattered memcpy introduces minor performance overhead compared to no memcpy (less than 8\% and 5\% for \allreduce and \alltoall, respectively). 
Kernel launching performs worse than kernel fusion for both small message latency and large message bandwidth. 
The suboptimal performance of Kernel launching is because of the high kernel launching overhead, especially for small messages. 

\subsubsection{PCIe overhead}
\begin{figure}[!t]
\centering
\includegraphics[width=0.3\textwidth]
{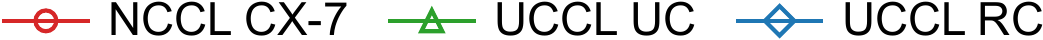}\vspace{0.05in}
\prefig
\begin{minipage}{0.23\textwidth}
    \centering
    \includegraphics[width=\linewidth]{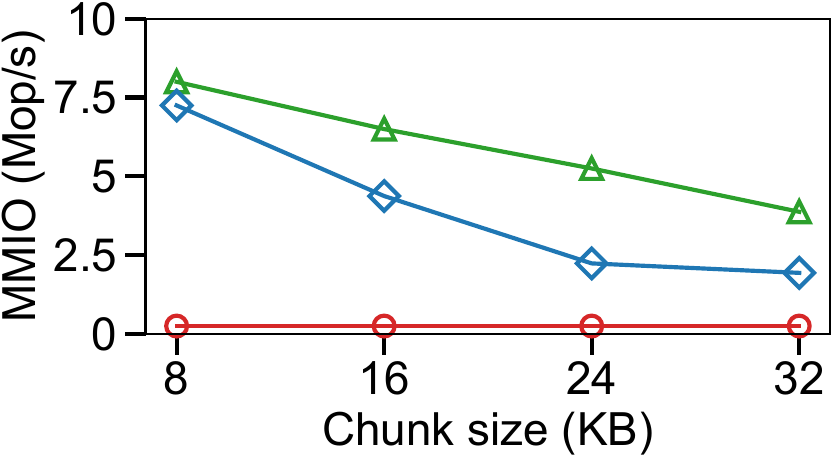}
    \label{fig:pcie_mmio_chunk}
\end{minipage}
\hfill
\begin{minipage}{0.23\textwidth}
    \centering
    \includegraphics[width=\linewidth]{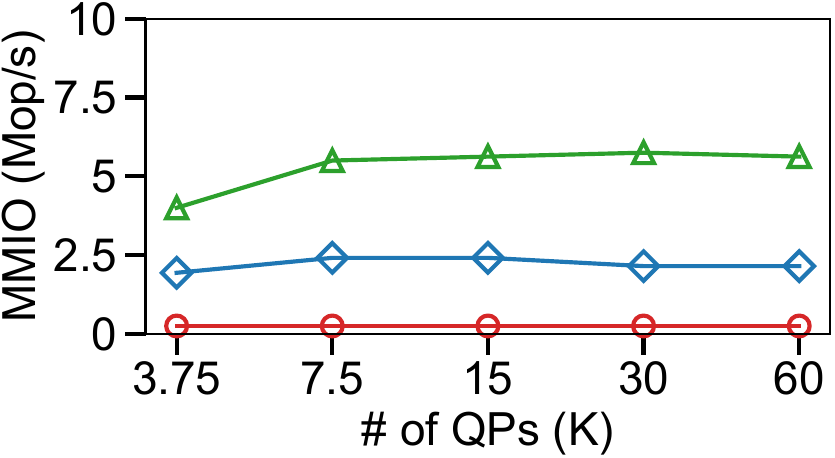}
    \label{fig:pcie_mmio_qp}
\end{minipage}
\postfigdouble
\caption{Number of MMIO events for each NIC under various chunk sizes and numbers of QPs on \cx.
}
\label{fig:pcie_mmio}
\postfigcaption\vspace{0.2in}
\end{figure}

\begin{figure}[!t]
\centering
\includegraphics[width=0.35\textwidth]
{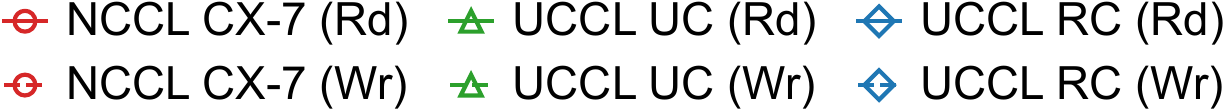}\vspace{0.05in}
\prefig
\begin{minipage}{0.23\textwidth}
    \centering
    \includegraphics[width=\linewidth]{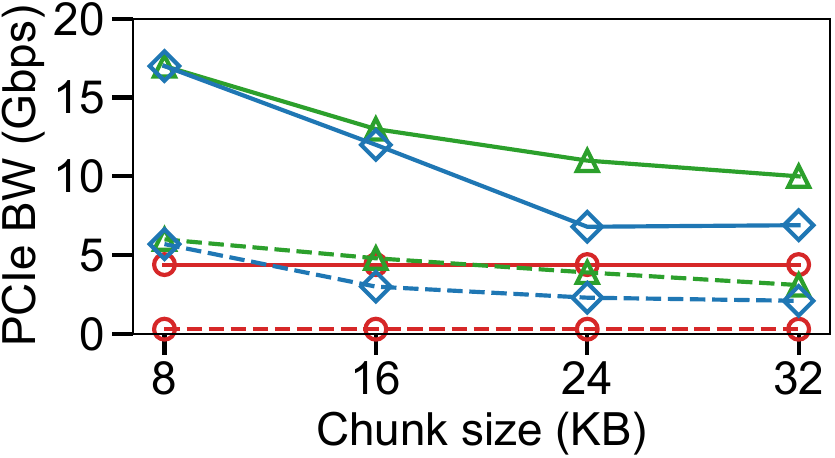}
    \label{fig:pcie_bw_chunk}
\end{minipage}
\hfill
\begin{minipage}{0.23\textwidth}
    \centering
    \includegraphics[width=\linewidth]{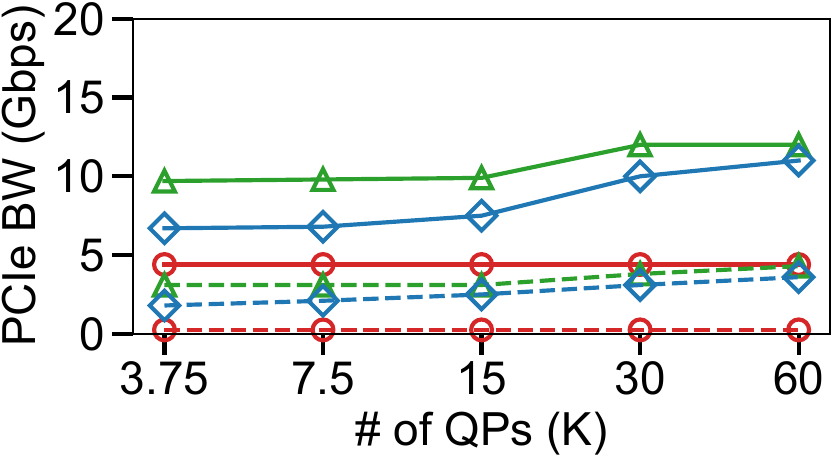}
    \label{fig:pcie_bw_qp}
\end{minipage}
\postfigdouble
\caption{Extra PCIe traffic for each NIC under various chunk sizes and numbers of QPs on \cx. Rd: PCIe device reads from CPU; Wr: PCIe device writes to CPU. 
}
\label{fig:pcie_bw}
\postfigcaption\vspace{0.1in}
\end{figure}

This experiment aims to study the PCIe overhead introduced by \sysname's multipath design. 
We measure the PCIe overhead on our \cx testbed, while the AWS virtualization layer prevents us from accessing low-level PCIe metrics. 
We rerun the \alltoall experiments in \S\ref{ssec:collective_perf} and quantify MMIO events and CPU-NIC PCIe traffic using \texttt{pcm-pcie}~\cite{pcm_pcie}. 
Figure~\ref{fig:pcie_mmio} and \ref{fig:pcie_bw} show the results. 
As expected, \sysname incurs higher MMIO activity and PCIe bandwidth consumption than the vanilla NCCL atop CX-7. 
\sysname incurs more MMIO events because it posts more verbs: \sysname transfers a small 32KB chunk per verb, while the vanilla NCCL atop CX-7 transfers a transport-buffer-sized message (\eg, default 128KB) per verb. 
The extra PCIe traffic of \sysname primarily comes from NIC swapping QPs (\ie, fetching uncached QP contexts from CPU memory) as \sysname uses 256 UC/RC QPs for multipath. 
\sysname UC consumes slightly more bandwidth than RC because it implements reliability and selective retransmission in software on the CPU. 

We conclude that the number of MMIO events and extra PCIe traffic increase significantly with smaller chunk sizes, but the number of QPs does not nearly affect them. 
\sysname makes a reasonable trade-off between the control decision granularity and performance based on this observation and uses a 32KB chunk size as default (\S\ref{ssec:efficient_sw}). 
We note that the extra PCIe bandwidth overhead is minor compared to the PCIe link capacity. For example, the capacity of PCIe 5.0 $\times$ 16 on \cx is 512 Gbps in each direction, and the overhead of \sysname (32KB, 60k QPs) accounts for less than 2.5\% (12/512=2.3\%).


\end{document}